\documentclass[11pt,a4paper]{article}

\usepackage{jheppub} 

\newcommand{\mre}{\mathrm{e}}
\newcommand{\mrd}{\mathrm{d}}

\newcommand{\bS}{\mathbf{S}}
\newcommand{\bSx}{\bS(x)}
\newcommand{\bSy}{\bS(y)}

\newcommand{\Z}{{\mathbb{Z}}}

\newcommand{\IDR}{\overline{I}}

\newcommand{\pz}{\partial_0}
\newcommand{\pmu}{\partial_\mu}
\newcommand{\pnu}{\partial_\nu}

\newcommand{\mcA}{\mathcal{A}} 
\newcommand{\mcF}{\mathcal{F}}

\newcommand{\mcW}{\mathcal{W}}
\newcommand{\mcWov}{\overline{\mathcal{W}}}
\newcommand{\Wov}{\overline{W}}

\newcommand{\vpi}{\vec{\pi}} 
\newcommand{\vpix}{\vec{\pi}(x)} 
\newcommand{\vpiy}{\vec{\pi}(y)} 
\newcommand{\vpiu}{\vec{\pi}(u)} 
\newcommand{\vpiv}{\vec{\pi}(v)}

\newcommand{\phat}{\hat{p}}
\newcommand{\psum}[1]{\sum_{#1}\rule{0pt}{2.5ex}'\;}
\newcommand{\psump}{\sum_{p}\rule{0pt}{2.5ex}'\;}

\newcommand{\order}[1]{\mathrm{O}\left( #1 \right) }

\newcommand{\bfp}{\mathbf{p}}
\newcommand{\bfz}{\mathbf{z}}

\newcommand{\Ls}{L_s}
\newcommand{\Lt}{L_t}
\newcommand{\Lhat}{\widehat{L}}
\newcommand{\ellhat}{\hat{\ell}}
\newcommand{\ds}{d_s}
\newcommand{\fm}{\mathrm{fm}}

\newcommand{\msbar}{{\rm \overline{MS\kern-0.14em}\kern0.14em}}
\newcommand{\phm}{\phantom{-}}

\title{Matching effective chiral Lagrangians with dimensional and 
lattice regularizations}

\author[a]{F.\ Niedermayer} \author[b]{and P.\ Weisz}

\affiliation[a]{Albert Einstein Center for Fundamental Physics, \\
  Institute for Theoretical Physics, University of Bern, 
  Switzerland} 
\affiliation[b]{Max-Planck-Institut f\"ur Physik, 80805
  Munich, Germany}

\emailAdd{niedermayer@itp.unibe.ch} \emailAdd{pew@mpp.mpg.de}

\abstract{We compute the free energy in the presence of a chemical potential
  coupled to a conserved charge in effective O($n$) scalar field theory
  (without explicit symmetry breaking terms) to third 
  order for asymmetric
  volumes in general $d$--dimensions, using dimensional (DR) and lattice
  regularizations.  This yields relations between the 4-derivative couplings
  appearing in the effective actions for the two regularizations, which in
  turn allows us to translate results, e.g.\ the mass gap in a finite periodic
  box in $d=3+1$ dimensions, from one regularization to the other. Consistency
  is found with a new direct computation of the mass gap using DR.  For the
  case $n=4, d=4$ the model is the low-energy effective theory of QCD with
  $N_\mathrm{f}=2$ massless quarks. The results can thus be used to obtain 
  estimates of low energy constants in the effective chiral Lagrangian 
  from measurements of the low energy observables, including the low lying 
  spectrum of $N_\mathrm{f}=2$ QCD in the $\delta$--regime using lattice 
  simulations, as proposed by Peter Hasenfratz, or from the susceptibility 
  corresponding to the chemical potential used.}

\dedicated{We dedicate this paper to the memory of Peter Hasenfratz,\\
 a great physicist, our good friend and colleague.}

\subheader{\hfill MPP-2015-266}

\begin{document}

\maketitle

\section{Introduction}
\label{Introduction}

There is now much confidence that Quantum Chromodynamics (QCD) 
is the theory of the strong interactions,
and there is good evidence that in this theory with massless quarks chiral
flavor symmetry is spontaneously broken. The low energy phenomena in systems
with spontaneously broken symmetry are governed by the dynamics of the
Goldstone bosons (pions in the case of QCD).  This can be described by an
effective field theory, and the calculations can be performed by chiral
perturbation theory $\chi$PT \cite{Wei79,Gas84}.

The interplay between $\chi$PT and QCD has been extremely fruitful. In early
times of lattice simulations of QCD when light dynamical quarks could not be
simulated efficiently, $\chi$PT was used to extrapolate the data to smaller
pion masses $m_\pi$. Lately, since simulations at physical pion masses became
feasible, one can use lattice data to obtain the parameters in the chiral
Lagrangian, the pion decay constant $F_\pi$ and the low energy constants
(LEC's), from the underlying microscopic theory QCD more precisely than from
phenomenology. For a detailed summary of 
various determinations of the LEC's the reader is 
referred to the FLAG review \cite{Aoki13}.

Both $\chi$PT computations and lattice simulations of QCD can be done in
special environments where physical experiments cannot be envisaged.  One can
study the dependence on parameters (such as quark masses), and one can place
the system into a space-time box of size $\Lt\times\Ls^{d-1}$ and study the
dependence of physical quantities on the box size $\Ls$ of the order a few
fermi. Leutwyler was the first to systematize the different regimes of QCD in
a finite box \cite{Leu87}.  One special environment is the so called
$\delta$--regime where the system is in a periodic spatial box of sides $\Ls$
and $m_\pi\Ls$ is small (i.e.\ small or zero quark mass) whereas $F_\pi\Ls$ is
large.

In 2009 Hasenfratz \cite{Has09} pointed out that promising observables in the
$\delta$-regime are the low lying stable masses.  Firstly measuring low lying
stable masses to good precision is among the easiest numerical tasks.
Secondly the finite box size introduces an infra-red cutoff which allows to
study the chiral limit in a first stage and switching on the symmetry breaking
terms later.

For massless two-flavor QCD the relevant $\chi$PT has
SU(2)$\times$SU(2)$\simeq O(4)$ symmetry.  It has been shown by Leutwyler
\cite{Leu87} that in the leading order of $\chi$PT, with general (unbroken)
O($n$) symmetry (and $d=4$), the spectrum is given by a quantum mechanical
rotator $E(l) = l(l + n - 2)/(2\Theta)\,,\,\,l = 0,1,2,\ldots$, the ``angular
momentum'' being the O($n$) isospin, with moment of inertia $\Theta=
F^2\Ls^3$ fixed by the decay constant $F$ (in the chiral limit).

The next-to-leading order (NLO) term of the expansion in $1/(F^2\Ls^2)$ has
been calculated in \cite{Has93}.  The level spectrum is to this order still
governed solely by $F$, so an evaluation of this spectrum on the lattice
potentially gives a good initial estimate of $F$.  Since the NLO correction
turned out to be large, however, it was important to calculate the 
next-to-next-to-leading order (NNLO) term;
furthermore the chiral logs and the LEC's $l_1,l_2$ (in the 4-derivative terms
in the effective action) first appear at this order.

Two independent results for the NNLO correction have been presented.  The
first is by Hasenfratz \cite{Has09} using dimensional regularization (DR).
His procedure, which was quite involved, was to consider a volume infinite in
the time direction, and to separate the degrees of freedom in the
$\delta$--regime into (spatially constant) slow and fast modes.  The latter
are then integrated out (treated in PT) resulting in an effective Lagrangian
for the slow modes, an O($n$) rotator with a modified moment of inertia, whose
energy excitations are much smaller than those of the standard Goldstone boson
excitations carrying finite momenta $\ge 2\pi/\Ls$.

The second computation was by Niedermayer and Weiermann \cite{Nie09} using
lattice regularization; it involved generalizing the computation of the
small-volume mass gap in the 2d O($n$) non-linear sigma-model by L\"{u}scher,
Weisz, and Wolff \cite{Lue91} to higher dimensions $d>2$.

Of course the physical content of a QFT is independent of regularization.  The
matching of UV regularizations of renormalizable asymptotically free theories
can be obtained by determining the ratio of $\Lambda$-parameters which just
involves a 1-loop calculation.  Here we have an effective QFT and the matching
of different regularizations in such theories is, as far as we know, still a
relatively untouched problem.  In particular the results of the two NNLO
computations referred to above could not be quantitatively compared, apart
from the chiral logs, since relations between the couplings of the
4-derivative terms in the effective Lagrangians were unknown.  In this paper
we have closed this gap.

Here we compute the change in the free energy due to a chemical potential
coupled to a conserved charge in the non-linear O($n$) sigma model with two
regularizations, lattice regularization (with standard action) in sect.~2 and
DR in sect.~3.  The computation is performed in a general $d$-dimensional
volume with periodic boundary conditions in all directions. The volume is left
asymmetric.  This freedom allows us for $d=4$ in sect.~4 to establish two
independent relations among the 4-derivative couplings appearing in the
effective actions of the two regularizations.  These relations in turn allow
us to convert results for physical quantities computed by the lattice
regularization to those involving scales introduced in DR.  Computations on
the lattice, although algebraically more involved than analogous continuum
computations, have the advantage that they are conceptually ``fool-proof''.
Computations with DR are however often tricky starting at two loops.

In particular one of the relations referred to above allow us to convert the
result of the mass gap computed on the lattice in \cite{Nie09} to DR (in
sect.~5). Unfortunately the outcome of this does not agree with the result of
Hasenfratz \cite{Has09}.  We thus recomputed the mass gap with DR and thereby
obtained a result in complete agreement with that translated from the lattice.
We are thus quite confident that it is correct.

The sums and integrals which appear in our computation, in particular the
two-loop massless sunset diagram, are treated in a separate accompanying paper
\cite{Nie15b}.

In this paper we do not consider explicit O($n$) symmetry breaking. In QCD the
effect of including a small explicit symmetry breaking (a small quark mass)
has been done to LO in \cite{Leu87}, and to NLO by Weingart
\cite{Wei06,Wei06a}. 
In a recent paper Matzelle and Tiburzi \cite{Mat15}
study the effect of small symmetry breaking in the QM rotator picture, 
and extend the results for small non-zero temperatures.

The proposal to use the mass gap in the delta regime to determine
the LEC's of QCD has its own advantages and disadvantages. 
As opposed to a similar problem in the $\epsilon$-regime one does not need 
the value of the (finite-volume, renormalized) condensate,
which is a non-trivial task. It is rather insensitive to the LEC's
$l_1$ and $l_2$, which also means that it is a convenient
quantity to determine $F$. A computational difficulty is that one needs
a largely elongated lattice in the time direction. To follow the decay 
of the correlation function one needs $m(L_s) L_t\gg 1$ which means
$L_t/L_s \gg F^2 L_s^2 > 1$. (The last condition should hold since we have 
an expansion in $1/(F^2 L_s^2)$).
Another problem -- which is, however, not specific for the $\delta$-regime 
-- is to use a chirally invariant fermion action satisfying 
the Ginsparg-Wilson relation, if one would like to avoid the extra tuning 
of the bare quark mass approaching the chiral limit.

There are other important physical systems where the order parameter of the
spontaneous symmetry breaking is an O($n$) vector.  In particular in condensed
matter physics, anti-ferromagnetic layers are described by O(3) for $d=3$.
Here the NNLO computation is complete to order $1/\Ls^2$; for details and
comparisons to experiment see ref.~\cite{Has93}.

\section{The free energy on the lattice}
\label{sect2}

In this section we work with a hyper-cubic $d$-dimensional lattice of volume
$V=\Lt\times\Ls^{\ds}\,,$\,\,\,$\ds=d-1$.  Define the aspect ratio
$\ell\equiv\Lt/\Ls$.

The dynamical variables are spins $S_a(x)\,,a=1,\dots,n$ of unit length
$\bSx^2=1$ with periodic boundary conditions in all directions
$\bS(x+\Lt\hat{0})=\bS(x)=\bS(x+\Ls\hat{k})\,,\,\,k=1,\dots,\ds\,$, where
$\hat{\mu}$ is the unit vector in the $\mu$--direction.  We will often set the
lattice spacing $a$ to 1 and will restore it again only in selected equations.

\subsection{The effective lattice action}
\label{eff_latt_action}

The effective lattice action $\mcA$ is a sum over terms
\begin{equation}
  \mcA = A_2 + A_4 +\dots\,,
\end{equation}
where the classical continuum limit of $A_{2r}$ has $2r$ derivatives.  In this
paper we will only consider the expansion up to and including four
derivatives.

For $A_2$ we take the standard lattice action:
\begin{equation}
  A_2 = \frac{1}{2 g_0^2} \sum_{x \mu} \pmu
  \bSx\cdot \pmu \bSx = -\frac{1}{g_0^2} \sum_{x \mu}
  \bSx\cdot\bS(x+\hat{\mu}) + \mathrm{const}\,,
  \label{A_latt} 
\end{equation}
where $g_0$ is the bare coupling and $\partial_\mu$ is the lattice difference
operator (we will also need the backward difference operator $\partial_\mu^*$)
\begin{align}
  \partial_\mu f(x) &= f(x+\hat{\mu}) - f(x)\,,
  \\
  \partial_\mu^* f(x) &= f(x) - f(x-\hat{\mu})\,.
\end{align}

The most general form of the four derivative terms is given by \cite{Has93}
\begin{equation}
  A_4=\sum_{i=1}^5 \frac{g_4^{(i)}}{4} \left[
    A_4^{(i)} -c^{(i)} \sum_{x \mu} \pmu \bSx \cdot \pmu \bSx \right] \,,
  \label{A4total} 
\end{equation}
where
\begin{align}
  &A_4^{(1)}= \sum_x \Box \bSx \cdot \Box \bSx\,, \label{A41}
  \\
  &A_4^{(2)}=\sum_{x\mu\nu} \left[\pmu\bSx \cdot \pmu\bSx\right]
  \left[\pnu\bSx \cdot \pnu\bSx\right]\,,  \label{A42}
  \\
  &A_4^{(3)}=\sum_{x\mu\nu} \left[\pmu\bSx \cdot \pnu\bSx\right]
  \left[\pmu\bSx \cdot \pnu\bSx\right]\,,  \label{A43}
  \\
  &A_4^{(4)}=A_4^{(4a)} - \frac{1}{d+2}\left(A_4^{(2)}+2A_4^{(3)}\right)\,,
  \label{A44}
  \\
  &A_4^{(5)}=A_4^{(5a)}-\frac{1}{d+2}\left(2A_4^{(5b)}+A_4^{(5c)}
  \right)\,, \label{A45}
\end{align}
and
\begin{align}
  &A_4^{(4a)}=\sum_{x\mu} (\pmu\bS_x \cdot
  \pmu \bS_x)^2\,, \label{A44a}
  \\
  &A_4^{(5a)}=\sum_{x\mu} \Box_\mu\bSx \cdot \Box_\mu\bSx, \label{A45a}
  \\
  &A_4^{(5b)}=A_4^{(1)}\,,\label{A45b}
  \\
  &A_4^{(5c)}=\sum_{x\mu\nu} \pmu\pmu\bSx \cdot \pnu\pnu\bSx\,,\label{A45c}
\end{align}
where
\begin{equation}
  \Box_\mu\equiv \pmu \pmu^*\,,\,\,\,\,\,\,\,\,\,\,\Box=\sum_\mu \Box_\mu\,.
\end{equation}
In \eqref{A4total} we subtract a term proportional to the leading action $A_2$
from each of the 4-derivative interactions.  The coefficients $c^{(i)}$ serve
to remove the power-like divergence $1/a^p$ from the contribution of the
corresponding operator, (note $c^{(4)}=0$).  The subtracted operators then
renormalize multiplicatively.

The set of five operators above is redundant\footnote{As explained in
  \cite{Has93}, changing the field variable
  \begin{equation}
    \bSx \to [\bSx+\alpha \Box \bSx] /
    |\bSx+\alpha \Box\bSx|
    \label{Salpha}
  \end{equation}
  the leading term of the effective action produces 4-derivative terms:
  \begin{equation} \label{Aalpha} \frac12 \sum_{x\mu} \pmu \bSx\cdot \pmu \bSx
    \to \frac12 \sum_{x\mu} \pmu \bSx\cdot \pmu \bSx - \alpha \left(
      A_4^{(1)}-A_4^{(2)} \right) + \ldots
  \end{equation}
  up to terms with higher derivatives.}.  One can use this observation to
eliminate, say, the coupling $g_4^{(1)}$ (as in \cite{Has93}), or
alternatively, to check that the final result for physical quantities depends
only on the sum of the couplings, $g_4^{(1)}+g_4^{(2)}$.

The total action including only terms to this order is given by
\begin{equation}
  \mcA=Z_4A_2+\sum_{i=1}^5\frac{g_4^{(i)}}{4}A_4^{(i)}\,,
\end{equation}
where
\begin{equation}
  Z_4\equiv 1-\frac12 g_0^2\sum_{i=1}^5 g_4^{(i)}c^{(i)}\,.
\end{equation}

\subsection{Perturbative expansion}

After separating the zero mode as in \cite{Has84} and changing to $\vpi$
variables (with $\sum_x\vpix=0$) according to $\bS = (g_0 \vpi, \sqrt{1-g_0^2
  \vpi^2})$ we have
\begin{equation}
  A_{2,\mathrm{eff}}[\vec{\pi}]=A_2[\vec{\pi}]
  +A_{2,\mathrm{measure}}[\vec{\pi}]+A_{2,\mathrm{zero}}[\vec{\pi}]\,, 
\end{equation}
with
\begin{align}
  &A_{2,\mathrm{measure}}[\vec{\pi}]=
  \sum_x \ln\left[ 1-g_0^2 \vpix^2\right]^\frac{1}{2}\,,\\
  &A_{2,\mathrm{zero}}[\vec{\pi}]= -n_1\ln \frac{1}{V}\sum_x \left[ 1-g_0^2
    \vpix^2\right]^\frac{1}{2} \,,
\end{align}
where
\begin{equation}
  n_1\equiv n-1\,.
\end{equation}

$A_{2,\mathrm{eff}}$ has a perturbative expansion
\begin{equation}
  A_{2,\mathrm{eff}} = A_{2,0} + g_0^2 A_{2,1} + g_0^4 A_{2,2} + \ldots
  \label{Aeff} 
\end{equation}
where (here we will need only $A_{2,0}$ and $A_{2,1}$):
\begin{align}
  A_{2,0} & = \frac12 \sum_x\pmu \vpi(x) \cdot\pmu\vpi(x)\,,\label{A20}
  \\
  A_{2,1} & = A_{2,1}^{(a)}+A_{2,1}^{(b)}\,, \label{A21}
  \\
  A_{2,1}^{(a)} & = -\frac12\left(1-\frac{n_1}{V} \right) \sum_x\vpi(x)^2\,,
  \\
  A_{2,1}^{(b)} & = \frac18\sum_x\pmu\left[\vpi(x)^2\right]
  \pmu\left[\vpi(x)^2\right]\,.
\end{align}

We expand the couplings of the 4-derivative terms according to
\begin{equation}
  g_4^{(i)}=\sum_{r=0}g_{4,r}^{(i)}g_0^{2r}\,.
\end{equation}
Noting that the coefficients $c^{(i)}$ defined in (\ref{A4total}) are of order
$g_0^2$:
\begin{equation}
  c^{(i)}=\overline{c}^{(i)} g_0^2 + \ldots
  \label{cbar}
\end{equation}
the renormalization constant $Z_4$ has a perturbative expansion
\begin{equation}
  Z_4=1+\sum_{r=2}Z_{4,r}g_0^{2r}\,,\,\,\,\,\,\,\,\,\,
  \,\,\,\left(Z_{4,1}=0\right)\,,
\end{equation}
with
\begin{equation}
  Z_{4,2} = -\frac12 \sum_{i=1}^5 g_{4,0}^{(i)}\overline{c}^{(i)}\,.
\end{equation}

The total effective action has a perturbative expansion of the form
\begin{equation}
  \mcA=\sum_{r=0}\mcA_r g_0^{2r}\,, 
\end{equation}
with
\begin{equation}
  \mcA_0=A_{2,0}\,,
\end{equation}
and
\begin{equation}
  \mcA_1=A_{2,1} 
  +\sum_{i=1}^5\frac{g_{4,0}^{(i)}}{4}A_{4,1}^{(i)}\,, 
\end{equation}
with
\begin{align}
  &A_{4,1}^{(1)}= \sum_x \Box \vpix \cdot \Box \vpix\,, \\
  &A_{4,1}^{(i)}=0\,,\,\,\,i=2,3,4\,,  \\
  &A_{4,1}^{(5)}=A_{4,1}^{(5a)}
  -\frac{1}{d+2}\left(2A_{4,1}^{(5b)}+A_{4,1}^{(5c)}\right)\,,
\end{align}
with
\begin{align}
  &A_{4,1}^{(5a)}=\sum_{x\mu} \Box_\mu \vpix \cdot \Box_\mu \vpix\,,  \\
  &A_{4,1}^{(5b)}=A_{4,1}^{(1)}\,,  \\
  &A_{4,1}^{(5c)}=\sum_{x\mu\nu} \pmu \pmu \vpix \cdot \pnu \pnu\vpix \,.
\end{align}

The perturbative coefficients of expectation values are sums of expectation
values of products of $\vpi$ fields with respect to the Gaussian free field
action $\mcA_0$ which are denoted by $\langle\dots\rangle_0$. In particular
the 2-point function is given by
\begin{equation}
  \langle\pi_a(x)\pi_b(y)\rangle_0=\delta_{ab}G(x-y)\,,
\end{equation} 
with the free propagator
\begin{equation}
  G(x)=\frac{1}{V}\psump \frac{\mre^{ipx}}{\phat^2}\,,\,\,\,\,\,\,\,
  \phat_\mu=2\sin\frac{p_\mu}{2}\,,
\end{equation}
where the sum is over momenta $p_0=2\pi n_0/\Lt\,,\,\,n_0=0,\dots,\Lt-1$ and
$p_k=2\pi n_k/\Ls\,,\,\,n_k=0,\dots,\Ls-1$ and the prime on the sum means that
$p=0$ is omitted.

\subsection{The chemical potential}
\label{chempotlatt}

The action considered in this paper has no explicit O($n$) (``isospin'')
symmetry breaking term. In an infinite volume the symmetry is 
broken  spontaneously, Goldstone bosons (massless pions for QCD) appear, 
and the ground state of the theory is a condensed state of the Goldstone bosons.
However, spontaneous isospin symmetry breaking is precluded in the present
finite volume setup.

Because of the symmetry the isospin charge and its component $Q_{12}$
are well defined, conserved quantities; their values
characterize a given eigenstate of the Hamiltonian (or transfer matrix). 
These numbers do not depend on the details of the regularization 
and renormalization.
One introduces the chemical potential by modifying the transfer 
matrix\footnote{For clarity we re-introduced here the lattice spacing.}
$\hat{T}(a) \to \hat{T}(a) \mre^{ah Q_{12}}$.
By this one only re-weights the states in the partition function, 
shifting their energy by $h Q_{12}$, hence the $h$-dependent part of 
the free energy should not depend on these details either. 
Hence by comparing this quantity in two different regularizations 
one can relate the corresponding couplings.

In terms of the action introducing the chemical potential is done 
by the replacement
\begin{equation}
  \bSx \cdot \bSy\to  [\bSx \cdot \bSy]_h\,,
\end{equation}
with
\begin{equation}
  \begin{aligned} 
    \left[ \bSx \cdot \bSy \right]_h &= \mre^{(y_0-x_0)h} S_-(x)S_+(y)
    +\mre^{-(y_0-x_0)h}S_+(x)S_-(y)+\sum_{a=3}^n S_a(x)S_a(y)\\
    &=\bS(x) \cdot \bS(y) + i \sinh\left((y_0-x_0) h\right) \; \left[S_1(x)
      S_2(y) - S_2(x) S_1(y)\right]
    \\
    &\quad + \{\cosh\left((y_0-x_0) h\right) -1\} \left[S_1(x) S_1(y)+ S_2(x)
      S_2(y)\right]\,,
  \end{aligned}
  \label{SxSy}
\end{equation}
where $S_\pm= \frac{1}{\sqrt{2}}(S_1 \pm i S_2)\,$. This gives an additional
$h$-dependent part $\mcA_h$ to the total action of the form
\begin{equation}
  \mcA_h=Z_4A_{2h}+\sum_{i=1}^5\frac{g_4^{(i)}}{4}A_{4h}^{(i)}\,.
\end{equation}
Further writing
\begin{align}
  &A_{2h}=ihB_2+h^2C_2+\dots\,,\\
  &A_{4h}^{(i)}=ihB_4^{(i)}+h^2 C_4^{(i)}+\dots
  \label{A4h}\,,
\end{align}
we have
\begin{align}
  B_2 & = -\frac{1}{g_0^2} \sum_x
  \left[S_1(x) S_2(x+\hat{0}) - S_2(x) S_1(x+\hat{0})\right]\,, \\
  C_2 & =-\frac{1}{2g_0^2}\sum_x \left[ S_1(x) S_1(x+\hat{0}) + S_2(x)
    S_2(x+\hat{0})\right]\,,
\end{align}
and the terms $B_4^{(i)}, C_4^{(i)}$ are given in Appendix~\ref{AppA}.

The $h$-dependent part of the free energy $f_h$ defined by:
\begin{equation}
  \mre^{- V f_h } = \langle \mre^{-\mcA_h}
  \rangle_\mcA\, = 1 - \langle \mcA_h \rangle_\mcA + \frac12 \langle \mcA_h^2
  \rangle_\mcA + \ldots
  \label{free_energy} 
\end{equation}
giving up to the order $h^2$:
\begin{equation}
  V  f_h  = \langle \mcA_h\rangle_\mcA-\frac12\langle \mcA_h^2\rangle_\mcA  
  +\frac12 \langle \mcA_h\rangle_\mcA^2+\dots 
\end{equation}
For finite volumes the $\lim_{h\to0}\left(f_h/h^2\right)$ exists; we define
the corresponding susceptibility $\chi$ by
\begin{equation}
  \chi\equiv -2\lim_{h\to0}\left(f_h/h^2\right)\,.
\end{equation}

Now
\begin{equation}
  \langle B_2\rangle_\mcA=0=\langle B_4^{(i)}\rangle_\mcA
  \,\,\,\,\,\,\forall i\,,
\end{equation}
so we have
\begin{equation}
  \chi =-2\sum_{s=1}^5F_s\,,
\end{equation}
with
\begin{align}
  F_1 & = Z_4\frac{1}{V}\langle C_2 \rangle_\mcA\,,
  \label{F1def}
  \\
  F_2 & = \frac12 Z_4^2\frac{1}{V}\langle B_2^2\rangle_\mcA\,,
  \label{F2def}
  \\
  F_3 & = \sum_{i=1}^5\frac{g_4^{(i)}}{4}\frac{1}{V} \langle
  C_4^{(i)}\rangle_\mcA\,,
  \label{F3def}
  \\
  F_4 & = Z_4\sum_{i=1}^5\frac{g_4^{(i)}}{4}\frac{1}{V}\langle
  B_2B_4^{(i)}\rangle_\mcA\,,
  \label{F4def}
  \\
  F_5 & = \frac12\sum_{i j}\frac{g_4^{(i)}}{4}\frac{g_4^{(j)}}{4}
  \frac{1}{V}\langle B_4^{(i)}B_4^{(j)}\rangle_\mcA\,.
  \label{F5def}
\end{align}

In the following subsections where we compute the contributions $F_s$ we will
use the fact that the total action $\mcA$ is invariant under global O$(n)$
transformations of the spins, so that the expectation value of any observable
is equal to the expectation value of its average over O$(n)$ rotations
$\Omega$:\footnote{Recall that we consider finite volume where no spontaneous
symmetry breaking occurs.} 
\begin{equation}
  \langle \mathcal{O}[S]\rangle_\mcA=
  \langle \left[\mathcal{O}[S]\right]_\Omega\rangle_\mcA\,.
\end{equation}
For arbitrary spins ${\bf a,b,c,d}$ we will use the following averages
\begin{equation}
  \left[ a_1 b_1 + a_2 b_2 \right]_\Omega =
  \frac{2}{n} (\mathbf{a}\cdot\mathbf{b}) \,,
  \label{a1b1} 
\end{equation}
and
\begin{equation}
  \left[ (a_1 b_2 - a_2 b_1)(c_1 d_2 - c_2 d_1)
  \right]_\Omega = \frac{2}{n(n-1)}\left[
    (\mathbf{a}\cdot\mathbf{c})(\mathbf{b}\cdot\mathbf{d})
    -(\mathbf{a}\cdot\mathbf{d})(\mathbf{b}\cdot\mathbf{c})\right] \,.
  \label{abcd} 
\end{equation}

\subsubsection{\boldmath Computation of $F_1$ up to $\order{g_0^2}$}

Averaging over the rotations (using (\ref{a1b1})) we have
\begin{equation}
  \frac{1}{V}\left[ C_2 \right]_\Omega = -\frac{1}{ng_0^2}
  +\frac{1}{2ng_0^2} U \,,
\end{equation}
where
\begin{equation}
  U = \frac{1}{V}\sum_x \pz \bSx \cdot \pz \bSx \,.
 \label{U} 
\end{equation}
This has the perturbative expansion
\begin{equation}
  U = g_0^2 U_1 + g_0^4 U_2 + \ldots
  \label{Ua} 
\end{equation}
with
\begin{equation}
  \begin{aligned} 
    &U_1 = \frac{1}{V}\sum_x \pz\vpix \cdot \pz\vpix\,, \\
    &U_2 = \frac{1}{4V}\sum_x \left[\pz\vpix^2\right]^2\,.
  \end{aligned}
  \label{U12}
\end{equation}

Expanding \eqref{F1def} in a perturbative series
\begin{equation}
  F_1=-\frac{1}{ng_0^2}+\sum_{r=0}^\infty F_{1,r}g_0^{2r}\,, 
\end{equation}
we have at leading orders
\begin{equation}
  F_{1,0} = \frac{1}{2n}\langle U_1\rangle_0\,,
  \label{F10} 
\end{equation}
and
\begin{equation}
  F_{1,1} =\frac{1}{2n}\Bigl[-2Z_{4,2} + \langle
  U_2 \rangle_0 -\langle U_1A_{2,1}\rangle_0^c
  -\sum_{i=1}^5\frac{g_{4,0}^{(i)}}{4}\langle
  U_1A_{4,1}^{(i)}\rangle_0^c\Bigr]\,,
  \label{F11} 
\end{equation}
where the superscript $\phantom{}^c$ in $\langle\dots\rangle_0^c$ means the
connected part.  The correlation functions appearing in (\ref{F10},\ref{F11})
are given in Appendix~\ref{AppC1} yielding
\begin{equation}
  F_{1,0} =\frac{n_1}{2n}I_{11}\,,
  \label{F10x} 
\end{equation}
and
\begin{equation}
  \begin{aligned}
    F_{1,1} & =\frac{n_1}{2n}\Bigl[-\frac{2}{n_1}Z_{4,2}
    +I_{11}\left\{I_{10}-\frac14 I_{11}\right\} -\mcF_1+\left(
      1-\frac{n_1}{V}\right)I_{21}
    \\
    &-\frac{g_{4,0}^{(1)}}{2}I_{01}
    -\frac{g_{4,0}^{(5)}}{2}\left\{J_{21}-\frac{1}{d+2}\left(2I_{01}
        +\mcF_4\right)\right\}\Bigr]\,. \label{F11p}
  \end{aligned}
\end{equation}
Here $I_{nm},J_{nm},\mcF_1,\mcF_4$ are momentum sums defined in equations
(\ref{Inm}), (\ref{Jnm}), (\ref{Lnm}), (\ref{cF1}), (\ref{cF4}) respectively.

\subsubsection{\boldmath Computation of $F_2$ up to $\order{g_0^2}$}

Averaging over the rotations one has, using (\ref{abcd}),
\begin{equation}
  \frac{1}{V}\left[ B_2^2\right]_\Omega = \frac{4}{n n_1 g_0^4} W \,,
  \label{Bsq}
\end{equation}
where $W$ is given by
\begin{equation}
  W = \frac{1}{V} \sum_{x y} \nabla_0 \bSx\cdot
  \nabla_0 \bSy \left[\bSx \cdot \bSy -1\right] \,,
  \label{W} 
\end{equation}
where $\nabla_0 = \frac12 (\pz + \pz^*)$ is the symmetric derivative.  $W$ has
a perturbative expansion
\begin{equation}
  W = g_0^4 W_2 + g_0^6 W_3 + \ldots
  \label{Wa} 
\end{equation}
with (to the order we need)
\begin{align}
  & W_2 = \frac{1}{V}\sum_{x y}
  \left[\nabla_0\vpix \cdot \nabla_0\vpiy\right]\vpix\cdot\vpiy\,, 
  \label{W2}
  \\
  & W_3 = \frac{1}{2 V}\sum_{x y}
  \left[\nabla_0\vpix\cdot\nabla_0\vpiy\right]\vpix^2\vpiy^2\,.
  \label{W3}
\end{align}
Expanding \eqref{F2def} in a perturbative series
\begin{equation}
  F_2=\sum_{r=0}^\infty F_{2,r}g_0^{2r}\,,
\end{equation}
we have at leading order
\begin{equation}
  F_{2,0} = \frac{2}{nn_1}\langle W_2\rangle_0\,,
  \label{F20} 
\end{equation}
and at next order
\begin{equation}
  F_{2,1}=\frac{2}{nn_1}\Bigl[\langle W_3\rangle_0
  -\langle W_2A_{2,1}\rangle_0^c -\sum_{i=1}^5\frac{g_{4,0}^{(i)}}{4}\langle
  W_2A_{4,1}^{(i)}\rangle_0^c \Bigr]\,.
  \label{F21} 
\end{equation}
The correlation functions appearing in (\ref{F20}) and (\ref{F21}) are
computed in Appendix~\ref{AppC2} yielding
\begin{equation}
  F_{2,0} =\frac{2(n_1-1)}{n}\left[I_{21}-\frac14 I_{22}\right]\,, 
\end{equation}
and
\begin{equation}
  \begin{aligned}
    F_{2,1}&=\frac{1}{n}W_{3c}+\frac{2(n_1-1)}{n}\Bigl[ W_{3a}-2\mcF_2+\mcF_3
    + 2\left(1-\frac{n_1}{V} \right)\left\{I_{31}-\frac14 I_{32}\right\}
    \\
    &-g_{4,0}^{(1)}\left(I_{11}-\frac14 I_{12}\right)
    -g_{4,0}^{(5)}\left(J_{31}-\frac14 J_{32}-\frac{1}{d+2}\left\{
        2I_{11}-\frac12 I_{12}+\mcF_5\right\}\right) \Bigr]\,,
  \end{aligned}
\end{equation}
where $\mcF_2,\mcF_3\mcF_5$ are defined in (\ref{cF2},\ref{cF3},\ref{cF5}),
and $W_{3a},W_{3c}$ are defined through
\begin{align}
  W_{3a} &= -\sum_x G(x)^2 \nabla_0^2 G(x)\,,
  \label{W3a}
  \\
  W_{3c} & = \sum_x \nabla_0 G(x) \left[ (\pz G(x))^2 - (\pz^* G(x))^2 \right]
  = - \frac16 \sum_x \left[ \Box_0 G(x) \right]^3 \,.
  \label{W3c}
\end{align}

\subsubsection{Summary}
\label{summary_lat}

The computation of the leading contributions to $F_3,F_4,F_5$ follows similar
steps as in the subsections above and details are presented in Appendices
\ref{AppC3}-\ref{AppC5}.  Summarizing our results so far, the susceptibility
with standard lattice regularization is given by
\begin{equation}
  \chi = \frac{2}{ng_0^2} \left( 1 +
    \overline{R}_1 g_0^2 + \overline{R}_2 g_0^4 +\dots \right)\,,
  \label{fh_hsq} 
\end{equation}
with
\begin{equation}
  \overline{R}_1=-\frac{n_1}{2}I_{11}
  -2(n_1-1)\left(I_{21}-\frac14 I_{22}\right)\,,
\end{equation}
and
\begin{equation}
  \overline{R}_2=\overline{R}_2^{(a)}+\overline{R}_2^{(b)}\,,
 \label{R2latt} 
\end{equation}
\begin{equation}
  \begin{aligned} 
    \overline{R}_2^{(a)}&= -\frac12 n_1\left[I_{11}\left\{I_{10}-\frac14
        I_{11}\right\} -\mcF_1+\left( 1-\frac{n_1}{V}\right)I_{21}\right]
    \\
    &-W_{3c}-2(n_1-1)\left[ W_{3a}-2\mcF_2+\mcF_3 + 2\left(1-\frac{n_1}{V}
      \right) \left\{I_{31}-\frac14 I_{32}\right\}\right]\,,
  \end{aligned}
  \label{R2alatt}
\end{equation}
and
\begin{equation}
  \overline{R}_2^{(b)}=\sum_{i=1}^5g_{4,0}^{(i)}G^{(i)}\,,
  \label{R2blatt}
\end{equation}
with
\begin{align}
  G^{(1)}&= -\frac12\overline{c}^{(1)}
  +2I_{11}-\frac12I_{12}+n_1\left[I_{00}-\frac14I_{01}\right]\,,
  \\
  G^{(2)}&= -\frac12\overline{c}^{(2)}+2I_{11}+n_1I_{00}\,,
  \\
  G^{(3)}&= -\frac12\overline{c}^{(3)}+I_{00}+(n_1+1)I_{11}\,,
  \\
  G^{(4)}&= -\frac12\overline{c}^{(4)}
  -\frac{(n_1+2)}{(d+2)}\left[I_{00}-dI_{11}\right]\,,
  \\
  G^{(5)}&= -\frac12\overline{c}^{(5)}
  -\frac{2}{(d+2)}\left\{2I_{11}-\frac12I_{12}
    +n_1\left[I_{00}-\frac14I_{01}\right]\right\}
  \nonumber\\
  &-\frac{n_1}{(d+2)}\left[-(d+1)\left\{3I_{11}-I_{12}\right\}+\mcF_6
    -\frac14(d+2)J_{21}+\frac14\mcF_4\right]
  \nonumber\\
  &+(n_1-1)\left\{2J_{31}-\frac12 J_{32}
    -\frac{1}{(d+2)}\left[2\mcF_5+(d+1)\left(4I_{22}-I_{23}\right)
      -2\mcF_7\right]\right\}\,.
\end{align}
A check of \eqref{R2alatt} for the special case of $n=2$ is given in
Appendix~\ref{AppD}.

\subsection{Renormalization on the lattice}

The renormalization procedure depends on the dimension; in the following we
will consider the cases $d=2,3,4$.

\subsubsection{\boldmath Case $d=2$}

For $d=2$ the theory is renormalizable so we can set the 4-derivative
couplings $g_4^{(i)}$ to zero.  As is well known the theory is asymptotically
free.

A renormalized ``minimal'' lattice coupling $g_\mathrm{latt}(\mu)$ is defined
through
\begin{equation}
  \frac{1}{g_0^2}=\frac{1}{g_\mathrm{latt}^2(\mu)}
  -b_0\ln(a\mu)-b_1\ln(a\mu)g_\mathrm{latt}^2(\mu)+\ldots
\end{equation}
where $b_0,b_1$ are the universal 1-, and 2-loop coefficients of the
$\beta$-function \cite{Bre76,Ami80}:
\begin{equation}
  b_0=\frac{n-2}{2\pi}\,,\,\,\,\,\,\,\,\,
  b_1=\frac{n-2}{4\pi^2}\,.
  \label{betad2} 
\end{equation}
In the continuum limit
\begin{align}
  I_{11}&=\frac12+\order{a^2}\,,
  \\
  I_{21}&=\frac{1}{4\pi}\ln(\Ls/a)+I_{21;0}(\ell)+\order{a^2}\,,
  \\
  I_{22}&=\frac12-\frac{1}{2\pi}+\order{a^2}\,.
\end{align}
The coefficients $I_{nm;r}$ appearing in the large $\Ls/a$ expansion of
$I_{nm}$ are considered in \cite{Nie15b}.

So
\begin{equation}
  \overline{R}_1=-b_0\ln(\Ls/a)+\overline{r}_1+\order{a^2}\,,
\end{equation}
with
\begin{equation}
  \overline{r}_1=-\frac14
  -2(n-2)\left[I_{21;0}(\ell)+\frac{1}{8\pi}\right]\,.
\end{equation}

Next
\begin{align}
  I_{10}&=\frac{1}{2\pi}\ln(\Ls/a)+I_{10;0}(\ell)+\order{a^2}\,,
  \\
  I_{32}&=\frac{3}{16\pi}\ln(\Ls/a)+I_{32;0}(\ell)+\order{a^2}\,,
  \\
  \frac{1}{V}I_{31}&=I_{31;-2}(\ell)+\order{a^2}\,,
  \\
  \mcF_1&=\frac{1}{2\pi}\ln(\Ls/a) +I_{21;0}(\ell)+\frac12
  I_{10;0}(\ell)-\frac18+\order{a^2}\,,
  \\
  \mcF_2-I_{31}&=\frac{1}{8\pi^2}\ln^2(\Ls/a)
  +\frac{1}{2\pi}\left\{I_{21;0}(\ell)+\frac12 I_{10;0}(\ell)
    +\frac{1}{8\pi}-\frac{11}{32}\right\}\ln(\Ls/a)
  \nonumber\\
  &+\mcF_{2;0}+\order{a^2}\,,
  \\
  \mcF_{2;0}&=\left[I_{10;0}(\ell)-\frac14\right]
  \left\{I_{21;0}(\ell)-\frac18+\frac{1}{8\pi}\right\}
  -I_{31;-2}(\ell)-\frac14 I_{32;0}(\ell)\,,
  \\
  W_{3a}&=\frac{1}{8\pi^2}\ln^2(\Ls/a)
  +W_{3a;0x}(\ell)\ln(\Ls/a)+W_{3a;0}(\ell)+\order{a^2}\,,
  \\
  W_{3c}&=\frac{1}{48}+\order{a^2}\,.
\end{align}
So
\begin{equation}
  \overline{R}_2^{(a)}=-b_1\ln(\Ls/a)+\overline{r}_2+\order{a^2}\,,
\end{equation}
with
\begin{equation}
  \begin{aligned}
    \overline{r}_2&=-\frac{5}{96}+4(n-2)^2I_{31;-2}(\ell)
    \\
    &-2(n-2)\left[W_{3a;0}(\ell)+\frac{1}{64}
      -2\left\{I_{10;0}(\ell)-I_{21;0}(\ell)-\frac18-\frac{1}{8\pi}\right\}
      \left\{I_{21;0}(\ell)-\frac18+\frac{1}{8\pi}\right\}\right]\,,
  \end{aligned}
  \label{r2bar}
\end{equation}
where we have used the relation
\begin{equation}
  W_{3a;0x}(\ell)=\frac{1}{2\pi}\left[I_{10;0}(\ell)
    -\frac14+\frac{1}{4\pi}\right]\,.
\end{equation}
We thus obtain in the continuum limit:
\begin{equation}
  \begin{aligned}
    \chi & =\frac{2}{ng_\mathrm{latt}^2(\mu)} \left\{1
      +\left[-b_0\ln(\mu\Ls)+\overline{r}_1\right] g_\mathrm{latt}^2 (\mu)
      +\left[-b_1\ln(\mu\Ls)+\overline{r}_2\right] g_\mathrm{latt}^4(\mu)
      +\ldots\right\}  \\
    & = \frac{2}{ng_\mathrm{latt}^2(1/L_s)} \left\{1 + \overline{r}_1 g_{\rm
        latt}^2(1/L_s) + \overline{r}_2 g_\mathrm{latt}^4(1/L_s) + \ldots\right\}
  \end{aligned}
\end{equation}
which is interpreted as an expansion in the running lattice coupling 
$g_\mathrm{latt}(1/\Ls)$, the expansion being sensible only for physically 
small box size $\Ls$.

\subsubsection{\boldmath Case $d=3$}

For $d=3$ we set $g_0^2=1/(\rho_0 a)$, where $\rho_0$ is the bare spin
stiffness, and define a renormalized coupling $\rho$ as in \cite{Nie09}
through
\begin{equation}
  \frac{1}{\rho_0}=\frac{1}{\rho}\left(1+\frac{{b}_1}{\rho a}+
    \frac{{b}_2}{\rho^2 a^2}+\ldots\right)\,.
\end{equation}
Then we have
\begin{equation}
  \chi = \frac{2\rho}{n} \left( 1 +
    \frac{1}{\rho a}\hat{R}_1 + \frac{1}{\rho^2 a^2}\hat{R}_2 +\dots\right)
  \label{fh_hsqFd3} 
\end{equation}
with
\begin{align}
  \hat{R}_1 &= \overline{R}_1 - b_1\,,
  \\
  \hat{R}_2 &= \overline{R}_2 - b_2 + b_1^2\,.
\end{align}
From \cite{Nie15b} for $d=3$ $\overline{R}_1$ has a large $\Ls/a$ expansion of
the form
\begin{equation}
  \overline{R}_1=-\frac16-(n-2)I_{10;0}-2(n-2)I_{21;1}(\ell)\frac{a}{\Ls}+\dots
  \label{ovR1_d3}
\end{equation}
with $I_{10;0}=0.252731009859$, where the large $\Ls/a$ expansion of
$X_A$ is given by $X_A=\sum_{r=r_0}X_{A;r}(a/\Ls)^r$\,.  So for
renormalization at leading order we need
\begin{equation}
  b_1=-\frac16-(n-2)I_{10;0}\,.
\end{equation}

After choosing the $\overline{c}^{(i)}$ appropriately the terms in $\hat{R}_2$
coming from $\overline{R}_2^{(b)}$ are of order $a^3/\Ls^3$ \footnote{The
  couplings of the 4-derivative interactions in $d=3$ have dimension in the
  continuum formulation.}, so the continuum limit is determined only by
$\overline{R}_2^{(a)}$

Further $\overline{R}_2^{(a)}$ has a large $\Ls/a$ expansion for $d=3$ of the
form
\begin{equation}
  \overline{R}_2^{(a)}=\overline{R}_{2;0}^{(a)}
  +\overline{R}_{2;1}^{(a)}\frac{a}{\Ls}
  +\overline{R}_{2;2}^{(a)}\frac{a^2}{\Ls^2}+\dots
  \label{b2b1_d3}
\end{equation}
so renormalization requires
\begin{equation}
  b_2-b_1^2=\overline{R}_{2;0}\,,
\end{equation}
which gives
\begin{equation}
  b_2=b_{20}+b_{21}n_1+b_{22}n_1^2\,,                          
\end{equation}
with coefficients independent of $\ell$\footnote{Note
  $4I_{21;0}-I_{22;0}=2I_{10;0}-1/d$\,.}
\begin{equation}
  \begin{split}
    &b_{20}=2W_{3a;0}-W_{3c;0}=0.0102138509611\,, \\
    &b_{21}=\frac{1}{72}-2W_{3a;0}-I_{10;0}^2=-0.0659002864141\,, \\
    &b_{22}=I_{10;0}^2=0.0638729633447\,.
  \end{split}
  \label{b_ren_d3}
\end{equation}
Further we need
\begin{equation}
  0=\overline{R}_{2;1}^{(a)}=2(n-2)\left[
    -W_{3a;1}+\left\{I_{10;0}-\frac16\right\}I_{10;1}\right]\,,
\end{equation}
which we have verified numerically to high precision for $\ell=1$ and $\ell=2$
\cite{Nie15b}.

So finally we have for $d=3$ in the continuum limit:
\begin{equation}
  \chi = \frac{2\rho}{n} \left( 1 -
    \frac{1}{\rho\Ls}2(n-2)I_{21;1}(\ell) +
    \frac{1}{\rho^2\Ls^2}\overline{R}_{2;2}^{(a)}(\ell) +\dots\right)
  \label{fh_hsqFd3f} 
\end{equation}
with
\begin{equation}
  \overline{R}_{2;2}^{(a)}=2(n-2)\left\{
    -W_{3a;2}+2I_{21;1}\left( 
      I_{10;1} - I_{21;1}\right)+\frac{2}{\ell}(n-2)I_{31;-1}
  \right\}\,.
\end{equation}
This result for the susceptibility agrees with (2.29) of \cite{Has93}%
\footnote{with the identification of the notation used there (on the lhs):
  $\beta_1 = - I_{10;1}\,, \beta_2 = I_{20;-1}\,, \tilde{\beta}_1 = -6
  I_{21;1}\,,\,\,\, \tilde{\beta}_2 = (12/\ell) I_{31;-1}\,, \,\,\,\, \psi =
  W_{3a;2}-2 I_{10;1}I_{21;1}$.}. e.g.:
\begin{equation}
  \overline{R}_{2;2}^{(a)}(\ell)=
  \begin{cases} 
    -0.00920015939 - 0.007071685928\,(n-2)\,,& \text{for }\ell=1\,,\\
    \phantom{-}0.01560323409 - 0.01338624986\,(n-2)\,,& \text{for }\ell=2\,.
  \end{cases}
\end{equation}

\subsubsection{\boldmath Case $d=4$}
\label{rend4}

For $d=4$ we set $g_0^2=1/(F_0^2 a^2)$ and define a renormalized coupling $F$
(the pion decay constant in chiral PT in the chiral limit) through
\begin{equation}
  \frac{1}{F_0^2} = \frac{1}{F^2}\left( 1 +
    \frac{b_1}{F^2 a^2} + \frac{b_2}{F^4 a^4} + \order{1/(F a)^6}\right)\,.
  \label{g0ren} 
\end{equation}

After renormalization we have
\begin{equation}
  \chi = \frac{2F^2}{n} \left( 1 + \frac{1}{F^2
      a^2}\hat{R}_1 + \frac{1}{F^4 a^4}\hat{R}_2 +\dots \right)\,,
  \label{fh_hsqF}
\end{equation}
with
\begin{align}
  \hat{R}_1 &= \overline{R}_1 - b_1\,,
  \\
  \hat{R}_2 &= \overline{R}_2 - b_2 + b_1^2\,.
\end{align}

To cancel the $1/a^2$ terms in $\hat{R}_1/a^2$ one should require
\begin{equation}
  \begin{split}
    b_1 & = -\frac18 - (n-2) I_{10;0}
    \\
    & = 0.029933390231060214084 - 0.15493339023106021408 \, n_1\,.
  \end{split}
  \label{b1}
\end{equation}
This agrees with the result in \cite{Nie09}.  

Next since $I_{11;2}=0=I_{22;2}$ after renormalization we obtain
\begin{equation}
  \lim_{a\to0}\left[a^{-2}\hat{R}_1\right]=-\frac{2(n-2)}{\Ls^2}I_{21;2}\,.
\end{equation}

$\hat{R}_2/a^4$ has divergent terms proportional to $1/a^4$, $1/a^2$ and
$\log(a)$.  First we recall that the subtraction coefficients $c^{(i)}$ are
used to cancel the leading, $1/a^4$ contributions of the corresponding
operators.  In leading order $\overline{c}^{(i)}$ (defined in \eqref{cbar}) is
fixed by requiring
\begin{equation}
  \lim_{a/\Ls\to0}G^{(i)}=0\,,
\end{equation}
where $G^{(i)}$ are the coefficients in (\ref{R2blatt}), which leads to
\begin{equation}
  \begin{aligned}
    \overline{c}^{(1)} & = n - I_{12;0} = n - 0.7066242375215119838793013966\,,\\
    \overline{c}^{(2)} & = 2\,n-1 \,, \\
    \overline{c}^{(3)} & = \frac12 \, n + 2 \,, \\
    \overline{c}^{(4)} & = 0 \,,\\
    \overline{c}^{(5)} &=
    -0.030936190551839592713\,n-0.032327591899970596813\,.
  \end{aligned}
\end{equation}
In \cite{Nie09} the overall sign of $\overline{c}^{(i)}$ was wrong and we also
disagree here with sign of the constant term in $\overline{c}^{(5)}$.

Demanding the absence of the $1/a^4$ singularity in $\hat{R}_2/a^4$ determines
the second order coefficient
\begin{equation}
  \begin{aligned}
    b_2 &= I_{10;0}^2\, (n-2)^2 + \left( I_{10;0}^2 +\frac
      {1}{128}-2\,W_{3a;0} \right)\, (n-2)
    +\frac {1}{128}-W_{3c;0} \\
    &=0.024004355408 \,(n-2)^2 + 0.028115270716 \,(n-2) + 0.005536500909\,.
  \end{aligned}
  \label{b2}
\end{equation}
This agrees with \cite{Nie09}.

With the values for $\overline{c}^{(i)}$ above
$G^{(i)}=\order{a^4/\Ls^4}\,\,\forall i\,$. It follows that the $1/a^2$
contribution to $\hat{R}_2/a^4$ has no more free parameters and should vanish
identically.  This requires the relation
\begin{equation}
  W_{3a;2} =  \left(I_{10;0}-\frac18 \right)I_{10;2}\,,
\end{equation}
which indeed holds numerically (see \cite{Nie15b}).

As for the renormalization of the 4-derivative couplings one has
\begin{equation}
  g_{4,0}^{(i)} = k^{(i)} \log(aM_i) \,, \quad  i=1,2,3\,,
  \label{g4ren}
\end{equation}
while for $i=4,5$ they are not renormalized to this order.

Moreover it is easy to check that after choosing the $\bar{c}^{(i)}$ as
above\footnote{Note $I_{12;4}=0$.}
\begin{equation}
  G^{(1)}=G^{(2)}=\frac{a^4}{\Ls^4}\left[2I_{11;2}-\frac{n_1}{\ell}\right]
  +\order{a^6/\Ls^6}\,,
\end{equation}
so that the part of $\hat{R}_2$ contributing in the continuum limit depends on
$g_{4,0}^{(1)},g_{4,0}^{(2)}$ only through their sum
$g_{4,0}^{(1)}+g_{4,0}^{(2)}$, consistent with our general argument on the
redundancy of the 4-derivative operators in subsection~\ref{eff_latt_action}.
In the following we shall use this redundancy to set $g_4^{(1)}=0$.

The cancellation of the $\ln(\Ls/a)$ terms requires a relation for the
coefficient of the $N^{-4}\ln N$ term in $W_{3a}$\,,
\begin{equation}
  W_{3a;4x} = \frac{1}{48\pi^2}\left(10I_{11;4}
    +\frac{1}{\ell}\right) \,,
  \label{w3ax} 
\end{equation}
which is satisfied numerically to high precision.  Then the coefficients of
the logarithmic terms of $g_{4,0}^{(2)}$ and $g_{4,0}^{(3)}$ are fixed as:
\begin{align}
  -4\pi^2 k^{(2)} & = w_1 = \frac{n}{2}-\frac53\,,\label{w1} \\
  -4\pi^2 k^{(3)} & = w_2 = \frac23\,,\label{w2}
\end{align}
agreeing with refs.~\cite{Nie09} and \cite{Gas84}.

Noting the relation
\begin{equation}
  W_{3c;4}  = -\frac18  \left(4\,I_{10;0} -1 \right) I_{11;4} \,,
\end{equation}
which is satisfied by the numerical values in \cite{Nie15b}, and inserting
eqs.~\eqref{ovg42} and \eqref{ovg43} one obtains the continuum limit of
$\hat{R}_2/a^4$:
\begin{equation}
  \Ls^4\lim_{a\to0}\frac{\hat{R}_2}{a^4}
  =\widehat{H}_0 -H_2\frac{w_1}{4\pi^2}\ln(M_2\Ls)
  -H_3\frac{w_2}{4\pi^2}\ln(M_3\Ls) +H_4 g_{4,0}^{(4)}+H_5 g_{4,0}^{(5)}\,,
  \label{R2res} 
\end{equation}
where
\begin{equation}
  \begin{split}
    H_2 &=
    -\frac{n-1}{\ell}+2I_{11;4} \,, \\
    H_3 &= -\frac{1}{\ell} + n\,I_{11;4} \,, \\
    H_4 &= \frac16 (n+1) \left(\frac{1}{\ell}+ 4\,I_{11;4} \right) \,, \\
    H_5 &= \frac{n-1}{2\ell} + (3\,n-4)\,I_{11;4}
    + 2 (n -2)\left(J_{31;4}-2I_{22;4}\right) \,,
  \end{split}
  \label{H2345}
\end{equation}
and
\begin{equation}
  \widehat{H}_0 =-(n-2)\widehat{w}+\widehat{w}'I_{11;4}
  +\widehat{w}''\frac{1}{\ell}\,,
\end{equation}
with
\begin{align}
  \widehat{w}&=-2\left(I_{10;0}-\frac18\right)I_{10;4}
  +2W_{3a;4}-4(I_{10;2}-I_{21;2})I_{21;2}\,, 
  \label{what}
  \\
  \widehat{w}'&=\frac23(n-2)(I_{33;0}-4\,I_{32;0}) +\frac23
  \left(n-\frac34\right)I_{10;0}-\frac{1}{48}(5n-1)\,, 
  \label{whatp}
  \\
  \widehat{w}''&=\frac{1}{24}(3\,n^2+n-12) I_{10;0}
  -(n-2)\left\{\left(n-\frac43 \right)I_{32;0} -\frac16 I_{33;0} +
    \frac{1}{24}\right\}+ 4(n-2)^2 I_{31;0}\,.
  \label{whatpp}
\end{align}

Note that the coefficient of $g_{4,0}^{(4)}$ in eq.~\eqref{R2res} vanishes for
the hyper-cubic case, $\ell=1$.

\section{The free energy with dimensional regularization}
\label{sect3}

In this section we work in a continuum volume $V=\Lt\times\Ls^{\ds}\,,
$\,\,\,$\ds=d-1$. Again the dynamical variables are spins
$S_a(x)\,,a=1,\dots,n$ of unit length $\bSx^2=1$ with periodic boundary
conditions in all directions.  We will dimensionally regularize by adding $q$
extra compact dimensions of size $\Lhat$ (also with pbc) and analytically
continue the resulting loop formulae to $q=-2\epsilon$. We define
$D=d+q\,,V_D=V\Lhat^q$\,.  We denote aspect ratio of the extra dimensions by
$\ellhat\equiv\Lhat/\Ls$. It is advantageous to treat these 
extra dimensions with a different size, since an extra check of the 
calculation is provided by the requirement that
physical quantities are independent of this choice.

Many of the formulae are similar to those with lattice regularization and we
will duplicate many of the notations hoping that this will not lead to
confusion.

\subsection{The effective action}
\label{ef_action}

The effective action $\mcA$ is a sum over terms
\begin{equation}
  \mcA = A_2 + A_4 +\dots\,,
\end{equation}
where $A_{2r}$ has $2r$ derivatives. $A_2$ is simply given by
\begin{equation}
  A_2 = \frac{1}{2 g_0^2} \int_x \sum_{\mu}
  \partial_\mu \bS(x)\cdot\partial_\mu \bS(x)\,. 
  \label{A_cont} 
\end{equation}

The four derivative terms are
\begin{equation}
  A_4=\sum_{i=2,3}\frac{g_4^{(i)}}{4}A_4^{(i)}\,,
  \label{A4total_cont}
\end{equation}
where (we use redundancy immediately here to set $g_4^{(1)}=0$),
\begin{align}
  &A_4^{(2)}=\int_x\sum_{\mu\nu} \left[\pmu\bSx \cdot \pmu\bSx\right]
  \left[\pnu\bSx \cdot \pnu\bSx\right]\,,
  \label{A42cont}\\
  &A_4^{(3)}=\int_x\sum_{\mu\nu} \left[\pmu\bSx \cdot \pnu\bSx\right]
  \left[\pmu\bSx \cdot \pnu\bSx\right]\,.
  \label{A43cont}
\end{align}

\subsection{Perturbative expansion}

After separating the zero mode and changing to $\vpi$ variables 
($\bS = (g_0 \vpi, \sqrt{1-g_0^2 \vpi^2})$)
\begin{equation}
  A_{2,\mathrm{eff}}[\vec{\pi}]=A_2[\vec{\pi}]
  +A_{2,\mathrm{zero}}[\vec{\pi}]\,.
\end{equation}
Note that the measure term is not present with dimensional regularization.
\begin{equation}
  A_{2,\mathrm{zero}}[\vec{\pi}]=
  -n_1\ln \left( 
    \frac{1}{V_D}\int_x \left( 1-g_0^2\vec{\pi}(x)^2\right)^\frac12 \right)\,.
\end{equation}
$A_{2,\mathrm{eff}}$ has a perturbative expansion
\begin{equation}
  A_{2,\mathrm{eff}} = A_{2,0} + g_0^2 A_{2,1} + \order{g_0^4}\,,
  \label{Aeff_cont} 
\end{equation}
where
\begin{align}
  A_{2,0} & = \frac12\int_x\partial_\mu\vpi(x)\cdot\partial_\mu\vpi(x)\,,
  \label{A20_cont}
  \\
  A_{2,1} & = A_{2,1}^{(a)}+A_{2,1}^{(b)}\,, \label{A21_cont}
  \\
  A_{2,1}^{(a)} & = \frac{n_1}{2V_D}\int_x\vpix^2\,,
  \\
  A_{2,1}^{(b)} & = \frac18\int_x\pmu\left[\vpi(x)^2\right]
  \pmu\left[\vpi(x)^2\right]\,.
\end{align}

The total effective action has a perturbative expansion of the form
\begin{equation}
  \mcA=\sum_{r=0}\mcA_r g_0^{2r}\,, 
\end{equation}
with
\begin{equation}
  \mcA_r=A_{2,r}\,,\,\,\,\,r=0,1\,,
\end{equation}
since
\begin{equation}
  A_4^{(i)}=\order{g_0^4}\,,\,\,\,i=2,3\,.
\end{equation}

The free 2-point function is given by
\begin{equation}
  \langle\pi_a(x)\pi_b(y)\rangle_0=\delta_{ab}G(x-y)\,,
\end{equation} 
with propagator
\begin{equation}
  G(x)=\frac{1}{V_D}\psump \frac{\mre^{ipx}}{p^2}\,,
\end{equation}
where the sum is over momenta $p_\mu=2\pi n_\mu/L_\mu\,,\,\,n_\mu\in\Z$ and
the prime on the sum means that $p=0$ is omitted.

\subsection{The chemical potential}
\label{chempotcont}

The chemical potential $h$ is introduced by the substitution:
\begin{equation}
  \partial_0 \to \partial_0 -h Q\,,
\end{equation}
where $(QS)_1=iS_2\,,\,\,(QS)_2=-iS_1\,,$ and $(QS)_a=0\,,a=3,\dots,n$.

This gives an additional $h$-dependent part $\mcA_h$ to the total action of
the form
\begin{equation}
  \mcA_h=A_{2h}+\sum_{i=2,3}\frac{g_4^{(i)}}{4}A_{4h}^{(i)}\,.
\end{equation}
Further writing
\begin{align}
  &A_{2h}=ihB_2+h^2C_2+\dots\,,\\
  &A_{4h}^{(i)}=ihB_4^{(i)}+h^2 C_4^{(i)}+\dots
  \label{A4h_DR}\,,
\end{align}
we have
\begin{align}
  B_2 & = -\frac{1}{g_0^2} \int_x\,j_0(x)\,,\,\,\,\,\,\,
  j_\mu(x)=S_2(x)\partial_\mu S_1(x)-S_1(x)\partial_\mu S_2(x)\,,\\
  C_2 & = \frac{1}{2g_0^2}\int_x [Q\bSx]^2\,.
\end{align}
For the operator 2:
\begin{align}
  B_4^{(2)}&=-4\int_x \partial_\mu\bSx\cdot\partial_\mu\bSx\,j_0(x)\,,
  \\
  C_4^{(2)}&=-2\int_x\left\{\partial_\mu\bSx\cdot\partial_\mu\bSx
    \left[S_1(x)^2+S_2(x)^2\right]+2\left[j_0(x)\right]^2\right\}\,,
\end{align}
and for the operator 3:
\begin{align}
  B_4^{(3)}&=-4\int_x\partial_0\bSx\cdot\partial_\mu\bSx\,j_\mu(x)\,,
  \\
  C_4^{(3)}&=-2\int_x\left\{\partial_0\bSx\cdot\partial_0\bSx
    \left[S_1(x)^2+S_2(x)^2\right]
    +2\left[j_0(x)\right]^2+\left[j_k(x)\right]^2\right\}\,.
\end{align}

The $h$-dependent part of the free energy $f_h$ is defined as in
(\ref{free_energy}).  Now
\begin{equation}
  \langle B_2\rangle_\mcA=0=\langle B_4^{(i)}\rangle_\mcA
  \,\,\,\,\,\,\forall i\,,
\end{equation}
so we have
\begin{equation}
  \chi =-2\sum_{s=1}^5F_s\,,
\end{equation}
with
\begin{align}
  F_1 & = \frac{1}{V_D}\langle C_2 \rangle_\mcA\,, \label{F1defcont}
  \\ 
  F_2  & = \frac12\frac{1}{V_D}\langle B_2^2\rangle_\mcA\,,  \label{F2defcont}
  \\
  F_3 & = \sum_{i=2,3}\frac{g_4^{(i)}}{4}\frac{1}{V_D} \langle
  C_4^{(i)}\rangle_\mcA\,,   \label{F3defcont} 
  \\ 
  F_4 & =
  \sum_{i=2,3}\frac{g_4^{(i)}}{4}\frac{1}{V_D}\langle
  B_2B_4^{(i)}\rangle_\mcA\,,   \label{F4defcont} 
  \\ 
  F_5 & = \frac12\sum_{i j}\frac{g_4^{(i)}}{4}
  \frac{g_4^{(j)}}{4} \frac{1}{V_D}\langle
  B_4^{(i)}B_4^{(j)}\rangle_\mcA\,.  \label{F5defcont} 
\end{align}

Averaging over the rotations we have simply
\begin{equation}
  \frac{1}{V_D}\left[ C_2 \right]_\Omega = -\frac{1}{ng_0^2}\,,
\end{equation}
and
\begin{equation}
  F_1=-\frac{1}{ng_0^2}\,.
\end{equation}
Next
\begin{equation}
  \frac{1}{V_D}\left[ B_2^2\right]_\Omega =
  \frac{4}{n n_1 g_0^4} W \,,
  \label{Bsq_cont} 
\end{equation}
with $W$ given by
\begin{equation}
  W = \frac{1}{V_D}\int_{x y}
  \partial_0 \bSx\cdot \partial_0 \bSy 
  \left[\bSx \cdot \bSy-1\right]  \,.
  \label{Wcont} 
\end{equation}
This has a perturbative expansion
\begin{equation}
  W = g_0^4 W_2 + g_0^6 W_3 + \ldots
  \label{Wa_cont}
\end{equation}
with
\begin{align}
  & W_2 = \frac{1}{V_D}\int_{xy}
  \left[\partial_0\vpix\cdot\partial_0\vpiy\right]\vpix\cdot\vpiy\,, 
  \label{W2_cont} 
  \\
  & W_3 = \frac{1}{2 V_D}\int_{xy}
  \left[\partial_0\vpix\cdot\partial_0\vpiy\right]\vpix^2\vpiy^2\,.
  \label{W3_cont}
\end{align}

Expanding \eqref{F2defcont} in a perturbative series
\begin{equation}
  F_2=\sum_{r=0}^\infty F_{2,r}g_0^{2r}\,,
\end{equation}
we have at leading order
\begin{equation}
  \begin{aligned}
    F_{2,0}&=\frac{2}{nn_1}\langle W_2\rangle_0 
    \\
    &=\frac{2(n-2)}{n}\int_x\left[\partial_0 G(x)\right]^2
    =\frac{2(n-2)}{n}\IDR_{21}\,,
  \end{aligned}
  \label{F20cont}
\end{equation}
where dimensionally regularized sums $\IDR_{nm}$ are formally defined by
\begin{equation}
  \IDR_{nm} = \frac{1}{V} {\sum_p}'
  \;\frac{\left(p_0^2\right)^m}{\left(p^2\right)^n} \,.
  \label{IDRnm} 
\end{equation}
Sums with $m=0$ were treated by Hasenfratz and Leutwyler \cite{Has90}; we
generalize their methods to sums with $m=1$ in \cite{Nie15b}.
 
At next order
\begin{equation}
  F_{2,1}=\frac{2}{nn_1}\left[\langle
    W_3\rangle_0 -\langle W_2A_{2,1}\rangle_0^c \right]\,.
  \label{F21cont} 
\end{equation}
First
\begin{equation}
  \langle W_3\rangle_0 = \frac{1}{2 V_D}\int_{xy}
  \langle \partial_0\vpix\cdot\partial_0\vpiy \vpix^2 \vpiy^2 \rangle_0 =
  n_1(n-2) \Wov\,,
  \label{W3av} 
\end{equation}
where
\begin{equation}
  \Wov = -\int_x G(x)^2\partial_0^2 G(x)\,.
  \label{W3a_DR}
\end{equation}
This 2-loop function, the ``massless sunset diagram'', is calculated 
in detail in \cite{Nie15b}.
 
Next
\begin{equation}
  \begin{aligned}
    \langle W_2 A_{2,1}^{(a)} \rangle_0^c &=\frac{n_1}{2V_D^2}\int_{xyu}
    \left\langle
      \partial_0\vpix\cdot\partial_0\vpiy (\vpix\cdot\vpiy)\vpiu^2
    \right\rangle_0^c
    \\
    &= \frac{2n_1^2(n-2)}{V_D^2}\int_{xyu}
    \partial_0^x \partial_0^y G(x-y)G(x-u)G(y-u)
    \\
    &= \frac{2n_1^2(n-2)}{V_D^2}{\sum_p}'\frac{p_0^2}{\left(p^2\right)^3} =
    \frac{2n_1^2(n-2)}{V_D}\IDR_{31}\,,
  \end{aligned}
  \label{W2A1aav}
\end{equation}
and
\begin{equation}
  \begin{aligned}
    \langle W_2 A_{2,1}^{(b)}\rangle_0^c &=
    \frac{1}{8V_D}\int_{xyu}\left[\partial_\mu^u\partial_\mu^v \left\langle
        \partial_0\vpix\cdot\partial_0\vpiy(\vpix\cdot\vpiy)
        \vpiu^2\vpiv^2\right\rangle_0^c\right]_{v=u}
    \\
    &= \frac{n_1(n-2)}{V_D}\int_{xyu} \partial_\mu^u \partial_\mu^v \Bigl[
    2\partial_0^x \partial_0^y G(x-y)G(x-u)G(y-v)G(u-v)
    \\
    &\left. -\partial_0^x G(x-u)\partial_0^yG(y-v)G(x-v)G(y-u) \Bigr]
    \right|_{v=u}
    \\
    &= n_1(n-2)\frac{1}{V_D^2}{\sum_{p q}}'
    \left[\frac{2p_0^2(p-q)^2}{\left(p^2\right)^3 q^2 } +\frac{p_0 q_0
        (p-q)^2}{\left(p^2\right)^2\left(q^2\right)^2}\right]
    \\
    &=
    2n_1(n-2)\left[\IDR_{21}\IDR_{10}+\IDR_{31}\IDR_{00}-\IDR_{21}^2\right]\,.
  \end{aligned}
  \label{W2A1bav}
\end{equation}
Note that $\IDR_{00}= -\Box G(0)= - 1/V_D$ since the dimensional
regularization sets $\delta(0)=0$.

For the contribution from the 4-derivative terms, averaging over rotations:
\begin{align}
  \left[C_4^{(2)}\right]_\Omega &=-\frac{4}{nn_1}\int_x
  \left\{n_1\partial_\mu\bSx\cdot\partial_\mu\bSx
    +2\partial_0\bSx\cdot\partial_0\bSx\right\}\,,
  \\
  \left[C_4^{(3)}\right]_\Omega &=-\frac{4}{nn_1}\int_x
  \left\{\partial_\mu\bSx\cdot\partial_\mu\bSx
    +n\partial_0\bSx\cdot\partial_0\bSx\right\}\,.
\end{align}
So to first order perturbation theory
\begin{equation}
  F_{3,1}=\frac{4}{n}\left\{
    \frac{g_4^{(2)}}{4}\left[\frac{n_1}{V_D}-2\IDR_{11}\right]
    +\frac{g_4^{(3)}}{4}\left[\frac{1}{V_D}-n\IDR_{11}\right]\right\}\,.
\end{equation}
Finally
\begin{equation}
  F_{4,1}=F_{5,1}=0\,.
\end{equation}

\subsection{Summary}
\label{summary_DR}

Summarizing the previous results,
the expansion of the susceptibility with DR is given by
\begin{equation}
  \chi = \frac{2}{n g_0^2} \left( 1 + g_0^2 R_1
    + g_0^4 R_2 + \ldots \right)\,,
  \label{chiDR} 
\end{equation}
with
\begin{equation}
  R_1 = -2(n-2)\IDR_{21}\,,
  \label{R1f} 
\end{equation}
and
\begin{equation}
  R_2=R_2^{(a)}+R_2^{(b)}\,,
\end{equation}
with
\begin{align}
  R_2^{(a)} &= 2(n-2)\left\{ -\Wov
    +2\IDR_{21}\left[\IDR_{10}-\IDR_{21}\right]
    +\frac{2(n-2)}{V_D}\IDR_{31}\right\}\,, 
  \label{R2af} 
  \\
  R_2^{(b)} &=-4\left\{ \frac{g_4^{(2)}}{4}\left[\frac{n_1}{V_D}-2\IDR_{11}\right]
    +\frac{g_4^{(3)}}{4}\left[\frac{1}{V_D}-n\IDR_{11}\right]\right\}\,.
  \label{R2bf} 
\end{align}

\subsection{\boldmath Case $n=2$}

Note that $R_1=0=R_2^{(a)}$ for $n=2$. This is easily seen since for this
special case the 2-derivative action with chemical potential is simply
\begin{equation}
  A = \frac{1}{2 g_0^2} \int_x \left(\partial_\mu
    \Phi(x) - i h \delta_{\mu 0}\right)^2 = \frac{1}{2 g_0^2} \int_{x}
  \left(\partial_\mu \Phi(x) \right)^2 -\frac{h^2}{2 g_0^2}V_D\,.
  \label{An2}
\end{equation}
Therefore there are no corrections to the leading term for the susceptibility
\begin{equation}
  \chi = \frac{1}{g_0^2}\,.
\end{equation}

\subsection{\boldmath Case $d=2$}

For $d=2$ the theory is renormalizable and as before we set the 4-derivative
couplings to zero.  Renormalization in the minimal subtraction (MS) scheme is
achieved by
\begin{equation}
  g_0^2=\mu^{2\epsilon}g_\mathrm{MS}^2Z_1\,,
\end{equation}
with
\begin{equation}
  Z_1^{-1}=1+\frac{b_0}{2\epsilon}g^2+\frac{b_1}{4\epsilon}g^4+\dots\,,
\end{equation}
where $b_0,b_1$ are as in \eqref{betad2}.

For $D\sim2$
\begin{equation}
  \IDR_{21}\sim -\frac{1}{4\pi}L^{-D+2}
  \left[\frac{1}{D-2}-\frac12\gamma_2+\kappa_{21}(D-2)+\dots\right]\,,
\end{equation}
where the functions $\gamma_i(\ell)$ are defined in \cite{Nie15b}.

Next
\begin{align}
  \IDR_{10}&=-\frac{1}{2\pi}L^{-D+2}
  \left[\frac{1}{D-2}-\frac12\alpha_1+\frac{1}{2\mathcal{V}}
    +\kappa_{10}(D-2)+\ldots\right]\,,
  \\
  \IDR_{31}&=\frac{L^2}{64\pi^2}\left[\gamma_3+1\right]\,,
  \\
  \Wov&=L^{-2D+4}\frac{1}{8\pi^2}\left[\frac{1}{(D-2)^2}
    +\frac{1}{(D-2)}\left(-\alpha_1-\frac12+\frac{1}{\ell}\right)
    +\overline{w}+\ldots\right]\,.
  \label{Wov_d2}
\end{align}
where $\alpha_i(\ell)$ are defined in \cite{Has90}.

In terms of the renormalized coupling
\begin{equation}
  \chi=\frac{2}{ng_\mathrm{MS}^2}\left\{1
    -b_0\left(\ln(\mu\Ls)+\frac12\gamma_2\right)g_\mathrm{MS}^2
    -b_1\left(\ln(\mu\Ls)+r_2\right)g_\mathrm{MS}^4
    +\ldots\right\}\,,
  \label{chi_d2}
\end{equation}
with
\begin{equation}
  r_2=\overline{w}-2\kappa_{10}-\frac12\gamma_2
  \left(\alpha_1-\frac{1}{\ell}-\frac12\gamma_2\right)
  -16\pi^2\frac{(n-2)}{V_D}\IDR_{31}\,.
  \label{r2} 
\end{equation}
 
For completeness we note that the free energy for large $h$ was computed
to NLO with DR at infinite volume in \cite{Has90a,Has90b} with the result
in the $\msbar$ scheme:
\begin{equation}
  \begin{aligned}
    f(h)-f(0)&=-\frac{h^2}{2}\Bigl[\frac{1}{g_{\msbar}^2(\mu)}
    -\frac{(n-2)}{2\pi}\left(\ln(\mu/h)+\frac12\right) +\order{g^2}\Bigr]
    \\
    &=-\frac{h^2}{2}\Bigl[\frac{1}{g_{\msbar}^2(h)}-
    \frac{(n-2)}{4\pi}+\order{g^2}\Bigr]\,.
  \end{aligned}
\end{equation}
Noting
\begin{equation}
  \frac{1}{g_{\msbar}^2(h)}=\frac{(n-2)}{2\pi}
  \Bigl[\ln(h/\Lambda_{\msbar})
  +\frac{1}{n-2}\ln\ln(h/\Lambda_{\msbar})
  +\dots\Bigr]\,,
\end{equation}
this result can be expressed as
\begin{equation}
  f(h)-f(0)=-\frac{(n-2)}{2\pi}\frac{h^2}{2}\Bigl[
  \ln\frac{h}{\Lambda_{\msbar}\sqrt{\mre}}
  +\frac{1}{(n-2)}\ln\ln(h/\Lambda_{\msbar})
  +\dots\Bigr]\,.
  \label{PTf}
\end{equation}
Eq.~\eqref{PTf} was compared to the result from a non-perturbative computation
invoking the Bethe ansatz \cite{Has90a,Has90b} thereby obtaining the exact
ratio of the mass gap to the $\Lambda$--parameter $m/\Lambda_\msbar$. Later
the thermodynamic Bethe ansatz equations were extended to study the spectrum
at finite volume \cite{Bal04}-\cite{Bal10}.

\subsection{\boldmath Case $d=3$}

For $d=3$ the contribution of the 4-derivative terms and are not relevant at
$\order{\Ls^{-2}}$ since $R_2^{(b)}=\order{\Ls^{-3}}$ \footnote{Note that for
  $d=3$ the couplings $g_4^{(i)}$ have dimension, in contrast to $d = 4$.}.
We remark however that because the theory is non-renormalizable, it is
expected that they are necessary to absorb divergences at higher orders.

For the sums contributing to $R_1,R^{(2a)}$ we have
\begin{align}
  \IDR_{10}&=-\beta_1 \Ls^{-1}\,,
  \label{IDR10_d3}
  \\
  \IDR_{21}&=\frac{1}{8\pi\Ls}\left(\gamma_2-2\right)\,,
  \\
  &=-\frac{1}{3\Ls}\beta_1\,\,{\rm for}\,\,\,\ell_1=\ell_2=\ell_3\,,
  \label{IDR21_d3}
  \\
  \IDR_{31}&=\frac{\Ls}{64\pi^2}\left(\gamma_3+2\right)\,,
  \label{IDR31_d3}
\end{align}
where the functions $\beta_i(\ell),\gamma_i(\ell)$ are defined in
\cite{Nie15b}. Also $\Wov$ has a finite limit for $D=3$, and the
results of numerical evaluation for $\ell=1,\ell=2$ are given in
\cite{Nie15b}.

The agreement of $R_1,R_2$ with the lattice results is evident for $d=3$
because of the direct relation of the DR sums to the associated coefficients
of the lattice sums:
\begin{equation}
  \IDR_{10}=I_{10;1}/\Ls\,,\,\,\,\,\IDR_{21}=I_{21;1}\Ls\,,\,\,\,\,
  \IDR_{31}=I_{31;-1}\Ls\,,\,\,\,\,\Wov=W_{3a;2}/\Ls^2\,.
  \label{Wov_d3}
\end{equation}

\subsection{\boldmath Case $d=4$}

In $d=4$ we set $g_0^2=1/F^2$ which is not renormalized with DR.

In ref.~\cite{Nie15b} we have computed the various functions appearing in
$R_1,R_2$.  First, for $d=4\,,$ $\IDR_{21}$ has a finite limit as $q\to0$:
\begin{equation}
  \IDR_{21}=\frac{1}{8\pi\Ls^2}\left(\gamma_2(\ell)-1\right)
  + \order{D-4}\,.
  \label{R1a}
\end{equation}

For $D\sim4$ we find for the 1-loop functions,
\begin{align}
  \IDR_{10}&= -\beta_1(\ell)\Ls^{-2}+\order{D-4}\,,
  \\
  \IDR_{31}&=\frac{1}{32\pi^2}\left[\ln\Ls-\frac{1}{D-4}
    +\frac12\gamma_3(\ell)\right]+\order{D-4}\,,
\end{align}
and for the 2-loop function
\begin{equation}
  \Wov=\frac{1}{16\pi^2\Ls^4}
  \left\{\left[\frac{1}{D-4}-2\ln\Ls\right]\mcW_0(\ell)
    +\frac{1}{3\ell}\ln(\ellhat)
    -\frac{10}{3}\mcW_1(\ell,\ellhat)
    +\mcWov(\ell)\right\}+\order{D-4}\,,
  \label{Wov_d4}
\end{equation}
with
\begin{equation}
  \mcW_0(\ell)=\frac53\left(\frac12-\gamma_1(\ell)\right)
  -\frac{1}{3\ell}\,.
\end{equation}
$\mcWov(\ell)$ is given in \cite{Nie15b}; also 
$\mcW_1(\ell,\ellhat)$ is given there albeit that explicit
expression is not needed here.

Putting the results together for $D\sim4$
\begin{equation}
  \begin{aligned}
    R_2^{(a)}&=2(n-2)\frac{1}{16\pi^2\Ls^4}\left\{
      -\left[\frac{1}{D-4}-2\ln\Ls\right]
      \left[\frac53\left(\frac12-\gamma_1(\ell)\right)
        +\frac{1}{\ell}\left(n-\frac73\right)\right]\right.
    \\
    &\quad +\frac{1}{\ell}\left(n-\frac73\right)\ln(\ellhat)
    +\frac{10}{3}\mcW_1(\ell,\ellhat)
    \\
    &\quad \left. -\frac12\left(\gamma_2(\ell)-1\right)^2
      -4\pi\beta_1\left(\gamma_2(\ell)-1\right)
      +\frac{1}{2\ell}(n-2)\gamma_3(\ell)-\mcWov(\ell)\right\}\,.
  \end{aligned}
  \label{R2a}
\end{equation}

For the 4-derivative terms we should identify
\begin{equation}
  \frac{g_4^{(2)}}{4}=-l_1\,,\,\,\,\,\
  \frac{g_4^{(3)}}{4}=-l_2\,,
\end{equation}
with the bare couplings $l_i$ of Gasser and Leutwyler \cite{Gas84} for the
standard $\overline{\mathrm{MS}}$ scheme:
\begin{equation}
  l_i=\frac{w_i}{16\pi^2}\left[
    \frac{1}{D-4}+\ln\left(\overline{c}\Lambda_i\right)\right]\,,
  \label{liGL}
\end{equation}
where $\ln\overline{c}=\overline{C}$ (defined in \eqref{ovC}), and
\begin{equation}
  w_1=\frac{n}{2}-\frac53\,,\quad w_2=\frac23 \,,
\end{equation}
are as given by \cite{Gas84} in (\ref{w1},\ref{w2})\footnote{In \cite{Gas84}
  only the $n=4$ result is given.  Often the notation $\gamma_i$ is used for
  $w_i$ above, but we have already used $\gamma_i$ in the context of 1-loop
  integrals.}.  In order to pick up all terms of $R_2^{(b)}$ finite in the
limit $D\to4$ we need also order $q=D-4$ terms of $\IDR_{11}$:
\begin{equation}
  \IDR_{11}=\frac{1}{\Ls^4}\left\{
    \frac12\left(1-q\ln\Ls\right)\left[\gamma_1(\ell)-\frac12\right]
    +q\mcW_1(\ell,\ellhat)\right\}+\order{q^2}\,.
\end{equation}
We then get for $D\sim4$:
\begin{equation}
  \begin{aligned}
    R_2^{(b)}&=\frac{1}{16\pi^2\Ls^4}\left\{
      2(n-2)\left[\frac{1}{D-4}-2\ln\Ls\right]
      \left[\frac53\left(\frac12-\gamma_1(\ell)\right)
        +\frac{1}{\ell}\left(n-\frac73\right)\right]\right.
    \\
    &-2(n-2)\left[\frac{1}{\ell}\left(n-\frac73\right)\ln(\ellhat)
      +\frac{10}{3}\mcW_1(\ell,\ellhat)\right]
    \\
    &\left.+4w_1\ln\left(\overline{c}\Lambda_1\Ls\right)
      \left[\frac{(n-1)}{\ell}-\gamma_1(\ell)+\frac12\right]
      +4w_2\ln\left(\overline{c}\Lambda_2\Ls\right)
      \left[\frac{1}{\ell}-\frac{n}{2}
        \left(\gamma_1(\ell)-\frac12\right)\right]\right\}\,.
  \end{aligned}
  \label{R2bDR}
\end{equation}
Summing the terms we have for $d=4$:
\begin{equation}
  \begin{aligned}
    R_2&= \frac{1}{16\pi^2\Ls^4}\left\{
      -2(n-2)\left[\frac12\left(\gamma_2-1\right)^2
        +4\pi\beta_1\left(\gamma_2-1\right)
        -\frac{1}{2\ell}(n-2)\gamma_3
        +\mcWov(\ell)\right]\right.
    \\
    &\left.+4w_1\ln\left(\overline{c}\Lambda_1\Ls\right)
      \left[\frac{(n-1)}{\ell}-\gamma_1+\frac12\right]
      +4w_2\ln\left(\overline{c}\Lambda_2\Ls\right)
      \left[\frac{1}{\ell}-\frac{n}{2}
        \left(\gamma_1-\frac12\right)\right]\right\}\,.
  \end{aligned}
  \label{R2tot}
\end{equation}
Here $\beta_1$ and $\gamma_i$ depend on $\ell=L_0/L_s$.

Note that not only do the poles at $D=4$ cancel, but also 
$\mcW_1(\ell,\ellhat)$, hence the physical
amplitude $R_2$ is independent of $\ellhat$, the aspect ratio of the extra
unphysical dimensions, as to be expected.

\section{\boldmath Matching the effective actions for $d=2$ and $d=4$}

\subsection{\boldmath Case $d=2$}

By matching the results for the susceptibility computed using lattice and
dimensional regularizations we should obtain the 2-loop relation between the
respective renormalized couplings
\begin{equation}
  g_\mathrm{latt}^2=g^2_\mathrm{MS}\left[1+X_1g^2_\mathrm{MS}+X_2g^4_\mathrm{MS}
    +\dots\right]\,.
\end{equation}
First noting
\begin{equation}
  I_{21;0}(\ell)=\frac{1}{8\pi}\left[\gamma_2+2\overline{C}+5\ln2\right]\,,
\end{equation}
where
\begin{equation}
  \overline{C}=-\frac12\left[\ln(4\pi)-\gamma_E+1\right]=-1.476904292\,,
  \label{ovC}
\end{equation}
at leading order we reproduce Parisi's result\footnote{converted from Pauli
  Villars regularization to DR} \cite{Par80}
\begin{equation}
  \begin{aligned}
    X_1&=\overline{r}_1+\frac12 b_0\gamma_2
    \\
    &=\frac{b_0}{2}\left[\ln\left(\frac{\pi}{8}\right)
      -\gamma_E\right]-\frac14\,.
  \end{aligned}
\end{equation}
The ratio of $\Lambda$ parameters is
\begin{equation}
  \frac{\Lambda_\mathrm{latt}}{\Lambda_\mathrm{MS}}
  =\exp\left(\frac{X_1}{b_0}\right)\,.
\end{equation}

At next order matching we get
\begin{equation}
  X_2-X_1^2=\overline{r}_2+b_1r_2\,.
\end{equation}
For our purposes it is sufficient to consider the case $\ell=\ellhat=1$ for
which
\begin{equation}
  \Wov=\frac{1}{D}\left[\IDR_{10}^2-\frac{1}{V_D}\IDR_{20}\right]\,,
  \,\,\,\,\,(\ell=\ellhat=1)\,,
\end{equation}
so that
\begin{equation}
  \overline{w}=2\kappa_{10}+\frac14\alpha_1^2-\frac18\left[\gamma_3+1\right]
  \,,\,\,\,\,\,(\ell=\ellhat=1)\,,
\end{equation}
and from \eqref{r2} (noting $\gamma_s= (2/d)(s-1) \alpha_{s-1}$ for $\ell=1$)

\begin{equation}
  \begin{aligned}
    r_2 &=\frac12\alpha_1-\frac14\alpha_2-\frac18
    -16\pi^2\frac{(n-2)}{V_D}\IDR_{31} \\
    &=-0.1022210828989128367197392 -16\pi^2\frac{(n-2)}{V_D}\IDR_{31}\,,
    \qquad \ell=1
  \end{aligned}
\end{equation}

On the lattice side we get for $\ell=1$ from \eqref{r2bar}
\begin{equation}
  \overline{r}_2  =-\frac{5}{96}
  -(n-2) \left(\frac{1}{2\pi} I_{10;0} - I_{20;-2} -\frac{1}{32}
    +\frac{1}{16\pi^2} + a_1\right)
  +4(n-2)^2I_{31;-2} 
\end{equation}
where we used
\begin{equation}
  W_{3a;0} = \frac12 I_{10;0}^2 - \frac12
  I_{10;0} I_{22;0} -\frac12 I_{20;-2}+\frac12 a_1\,,\qquad \ell=1
  \label{W3a_a1} 
\end{equation}
and $a_1$ is the infinite-volume quantity
\begin{equation}
  \begin{aligned}
    a_1 & = -\frac14 \int_{k,l} \frac{E_{k+l}-E_k-E_l}{E_k E_l E_{k+l}^2}
    \sum_\mu \widehat{(k+l)}^4_\mu \\
    & = -\frac12 \sum_x \left(G(x)-G(0)\right)^2 \Box_0^2 G(x) =
    0.0461636292439177762(1)
  \end{aligned}
  \label{a1def}
\end{equation}
(with $E_k=\hat{k}^2$).  Inserting the numerical values one gets
\begin{equation}
  \overline{r}_2=-\frac{5}{96}
  -(n-2)\, 0.02514054820286075900(1)
  +4(n-2)^2I_{31;-2}\,,\,\,\,\,\,\ell=1\,.
\end{equation}

Noting
\begin{equation}
  I_{31;-2}(\ell)=\frac{1}{\Ls^2\ell}\IDR_{31}\,,
\end{equation}
we obtain
\begin{equation}
  X_2-X_1^2=-\frac{5}{96} - 1.0947301436539277\,b_1\,.
  \label{X2mX1aq} 
\end{equation}

$X_2$ was first computed by Falcioni and Treves \cite{Fal86}:
\begin{equation}
  X_2-X_1^2=-\frac{5}{96}
  +b_1\left[h_1-\frac14+\frac12\ln\left(\frac{\pi}{8}\right)
    -\frac12\gamma_E\right]\,,
  \label{FalTre}
\end{equation}
with the value of $h_1$ given in \cite{Lue91}%
\footnote{$h_1=1/2-4\pi^2(a_1-1/32)$ with $a_1$ given in \eqref{a1def}.  The
  value of $h_1$ given in \cite{Fal86} was not very precise.}
\begin{equation}
  h_1=-0.088766484(1)\,,
\end{equation}
giving
\begin{equation}
  X_2-X_1^2=-\frac{5}{96}-1.094730144(1)\,b_1\,.
\end{equation}
The perfect agreement of our result \eqref{X2mX1aq} with the result obtained
above by an independent method gives an additional check on our formulae in
subsections~\ref{summary_lat}, \ref{summary_DR} which are valid for arbitrary
$d\ge 2$.

\subsection{\boldmath Case $d=4$}

The equality of the lattice and DR results for $d=4$ at sub-leading order one
requires
\begin{equation}
  I_{21;2} = \frac{1}{8\pi}(\gamma_2-1)\,,
  \label{I21g} 
\end{equation}
which we have proven in \cite{Nie15b}.

Comparing \eqref{R2latt}-\eqref{R2blatt} with \eqref{R2tot}, the coefficients
of $\ln(\Ls)$ agree due to the relation (see \cite{Nie15b}):
\begin{equation}
  I_{11;4} = \frac12 \left( \gamma_1-\frac12\right) \,.
  \label{I11g} 
\end{equation}
For general $n$ the matching equation has the form
\begin{equation}
  H_2 \,\overline{g}_{4,0}^{(2)} + H_3 \,\overline{g}_{4,0}^{(3)} 
  +H_4\, g_{4,0}^{(4)} + H_5\, g_{4,0}^{(5)} + H_0 = 0\,,
  \label{Heq} 
\end{equation}
where
\begin{align}
  \overline{g}_{4,0}^{(2)} &=g_{4,0}^{(2)}
  +\frac{1}{4\pi^2}w_1\ln(a\bar{c}\Lambda_1)=
  -\frac{1}{4\pi^2}w_1\ln\left(\frac{M_2}{\overline{c}\Lambda_1}\right)\,,
  \label{ovg42}
  \\
  \overline{g}_{4,0}^{(3)} &=g_{4,0}^{(3)}
  +\frac{1}{4\pi^2}w_2\ln(a\bar{c}\Lambda_2)= -\frac{1}{4\pi^2}w_2
  \ln\left(\frac{M_3}{\overline{c}\Lambda_2}\right)\,,
  \label{ovg43}
\end{align}
and
\begin{equation}
  H_0 = \widehat{H}_0+2(n-2)\left[
    \frac{1}{16\pi^2}\mcWov-2(I_{10;2}-I_{21;2})I_{21;2}\right]
  -\frac{(n-2)^2}{16\pi^2\ell}\gamma_3\,,
  \label{H0} 
\end{equation}
where we have used another identity:
\begin{equation}
  I_{10;2} = -\beta_1 \,.
  \label{I10b} 
\end{equation}
So we have
\begin{equation}
  H_0 =-(n-2)w+w'I_{11;4}+w''\frac{1}{\ell}\,,
\end{equation}
with
\begin{align}
  w&=2W_{3a;4}-\frac{1}{8\pi^2}\mcWov
  -\left(2I_{10;0}-\frac14\right)I_{10;4}\,,\label{wfn}
  \\
  w'&=\widehat{w}'\,,
  \\
  w''&=\frac{1}{24}(3\,n^2+n-12) I_{10;0} -(n-2)\left\{\left(n-\frac43
    \right)I_{32;0} -\frac16 I_{33;0} + \frac{1}{24}\right\}+ 4(n-2)^2
  i_{31;0}\,,
\end{align}
where $\widehat{w}'$ is given in (\ref{whatp}) and $i_{31;0}$ is defined by
\begin{equation}
  i_{31;0}=I_{31;0}(\ell)-\frac{1}{64\pi^2}\gamma_3(\ell) \,.
\end{equation}
Now we find numerically
\begin{equation}
  i_{31;0}=0.00211856418663447748445\,,\,\,\,\,\text{independent of}
  \,\,\,\ell\,,
  \label{I31g}
\end{equation}
so that $w''$ is independent of $\ell$ (as is also $w'$).

Now the coefficients $H_2,H_3,H_4,H_5$ in (\ref{Heq}) only involve the three
linearly independent $\ell$--dependent functions $I_{00;4}=-1/\ell$,
$I_{11;4}$ and $J_{31;4}-2I_{22;4}$ so that for consistency a relation for $w$
in (\ref{wfn}) of the form
\begin{equation}
  w =\frac{d_1}{\ell} + d_2 I_{11;4} + d_3 \left(J_{31;4}-2I_{22;4}\right) 
\end{equation}
should hold with some $\ell$-independent constants $d_1$, $d_2$, $d_3$.  From
numerical data sets with $\ell=1,2,3$ one finds $d_1=-0.00472740$, $d_2=
0.00026214$ and $d_3= 0.00000028$.  Inserting these values into the relation
with $\ell=4$ we indeed find consistency within the numerical errors (with the
difference in the 6th significant digit).  We will assume that actually
$d_3=0$, and with this one obtains from $\ell=1,2$ the values
$d_1=-0.00472752$, $d_2= 0.00026215$.

Let us define
\begin{align}
  G_1 &\equiv-n_1\overline{g}_{4,0}^{(2)}-\overline{g}_{4,0}^{(3)} 
  +\frac16 (n_1+2)g_{4,0}^{(4)}+\frac12\,n_1g_{4,0}^{(5)}\,, 
  \label{G1def} 
  \\
  G_2 &\equiv 2\overline{g}_{4,0}^{(2)}+(n_1+1)\overline{g}_{4,0}^{(3)}
  +\frac23(n_1+2)g_{4,0}^{(4)}+(3\,n_1-1)g_{4,0}^{(5)}\,.
  \label{G2def} 
\end{align}
Then matching requires
\begin{align}
  &0=G_1+q_0^{(1)}+(n-2) q_1^{(1)}+(n-2)^2 q_2^{(1)}\,, \label{G1eq2}
  \\
  &0=G_2+q_0^{(2)}+(n-2)q_1^{(2)}\,, \label{G2eq2}
  \\
  &0= 2 g_{4,0}^{(5)} - d_3 \,, \label{G3eq2}
\end{align}
with
\begin{align}
  q_0^{(1)}&=\frac{1}{12}\,I_{10;0} \,,
  \\
  q_1^{(1)}&=-\frac{1}{24}+\frac{13}{24}\,I_{10;0}
  -\frac23\,I_{32;0}+\frac16\,I_{33;0} - d_1\,,
  \\
  q_2^{(1)}&=\frac18 I_{10;0} - I_{32;0} +4 i_{31;0}\,,
  \\
  q_0^{(2)}&=-\frac{3}{16} + \frac{5}{6}\,I_{10;0}\,,
  \\
  q_1^{(2)}&=-\frac{5}{48}+\frac23\,I_{10;0}
  -\frac83\,I_{32;0}+\frac23\,I_{33;0} - d_2\,.
\end{align}
The numerical values are
\begin{align}
  q_0^{(1)}&=\phantom{-}0.0129111158 \,,
  \\
  q_1^{(1)}&=\phantom{-}0.0434608716 \,,
  \\
  q_2^{(1)}&=\phantom{-}0.011640543735\,,
  \\
  q_0^{(2)}&=-0.0583888414 \,,  \label{q02val}
  \\ 
  q_1^{(2)}&=-0.0152288420 \,.  \label{q12val}
\end{align}

For the special case $n=2$ the solution is:
\begin{align}
  g_{4,0}^{(4)}+g_{4,0}^{(5)} &= \frac{1}{16}
  -\frac13 I_{10;0}\,, \quad (n=2) \,, \label{g44g45}
  \\
  \overline{g}_{4,0}^{(2)}+\overline{g}_{4,0}^{(3)} &= \frac{1}{32} -
  \frac{1}{12}\,I_{10;0}\,,\quad (n=2) \,.  \label{g42g43}
\end{align}
Note in the continuum limit (e.g.\ for DR) $A_4^{(2)}=A_4^{(3)}$ for $n=2$.

\section{The mass gap}

The mass of the O($n$) vector particle in a periodic spatial volume
$\Ls^{d-1}$ was computed with lattice regularization for arbitrary $d$ in
ref.~\cite{Nie09} up to second order in perturbation theory.  It takes the
form
\begin{equation}
  m_1=\frac{n_1 g_0^2a^{d-2}}{2\Ls^{d-1}}\left\{1+g_0^2c_2(a/\Ls)
    +g_0^4\left[c_3(a/\Ls)+\sum_{j=2}^5g_{4,0}^{(j)}d_3^{(j)}(a/\Ls)\right]
    +\order{g_0^6}\right\}\,,
\end{equation}
where the coefficients $c_2(a/\Ls),c_3(a/\Ls),d_3(a/\Ls)$ \footnote{keeping
  the notation of \cite{Nie09} and not to be confused with previously
  mentioned quantities with the same letters!}  are given in appendix~B of
\cite{Nie09}; they depend on $d$, and for the case $d=2$ the coefficients
$c_2,c_3$ agree with those previously computed in \cite{Lue91}.

Here we will only discuss the case $d=4$. Results are often quoted in terms of
the moment of inertia $\Theta$ which is simply related to the mass gap through
\begin{equation}
  m_1=\frac{(n-1)}{2\Theta}\,.
\end{equation}
$\Theta$ has a perturbative expansion of the form
\begin{equation}
  \frac{\Theta}{F^2\Ls^3}=1+\Theta_1(F\Ls)^{-2}+\Theta_2(F\Ls)^{-4}+\dots
  \label{Theta_exp}
\end{equation}

After renormalization of the couplings as in subsect.~\ref{rend4}, the moment
of inertia in the continuum limit is given by \eqref{Theta_exp} with
coefficients determined from the lattice computation taken from eq.~(6.20) of
\cite{Nie09}
\begin{equation}
  \Theta_1^{\mathrm{latt}} =
  0.225784959441\,(n-2)\,,
  \label{R1mLATT} 
\end{equation} 
and
\begin{equation}
  \begin{split}
    \Theta_2^{\mathrm{latt}} = & -\frac{0.8375369106}{12\pi^2}
    \left[ (3\,n-10) \ln(M_2\Ls) + 2\,n \ln(M_3\Ls)\right] \\
    & + 0.55835794046(n+1) g_{4,0}^{(4)} \\
    & + (1.11639602502 \, n - 0.55771822866) g_{4,0}^{(5)}  \\
    & - 0.0489028095 + 0.0101978424 \, (n-2)\,.
  \end{split}
  \label{R2mLATT}
\end{equation}
Using the definitions in \eqref{ovg42},\eqref{ovg43} and \eqref{G2def} we can
rewrite this involving DR scales:
\begin{equation}
  \Theta_2^{\mathrm{latt}} =
  \overline{\theta}_2 -\frac{0.8375369106}{12\pi^2} \left[ (3\,n-10)
    \ln(\overline{c}\Lambda_1\Ls) + 2\,n
    \ln(\overline{c}\Lambda_2\Ls)\right]\,,
  \label{R2mLATT_DR} 
\end{equation}
with
\begin{equation}
  \begin{aligned}
    \overline{\theta}_2 = &
    0.8375369106 \, G_2 -1.396214707(n-2) g_{4,0}^{(5)} \\
    & - 0.0489028095 + 0.0101978424 \, (n-2)\,.
  \end{aligned}
  \label{ovtheta2}
\end{equation}
Finally using eq.~\eqref{G2eq2} (with \eqref{q02val},\eqref{q12val}) which was
obtained by matching lattice and DR results for the free energy, and assuming
$g_{4,0}^{(5)}=0$, we obtain
\begin{equation}
  \overline{\theta}_2 = 0.0229525597 \, (n-2)\,.
  \label{R2mLATT_DR1} 
\end{equation}
Note that the $(n-2)^0$ terms in $\overline{\theta}_2$ cancel to our numerical
precision of 10 digits.

The continuum limit of $\Theta_i$ should of course be regularization
independent. Unfortunately \eqref{R2mLATT_DR1} does not agree with the result
for the moment of inertia previously computed by Hasenfratz \cite{Has09} using
dimensional regularization.  For this reason we recomputed the mass gap with
DR using free boundary conditions in the time direction in an analogous way to
that used for the lattice computation.  The computation is rather lengthy and
here we only present the final result (for arbitrary $d$):
\begin{equation}
  m_1=\frac{n_1g_0^2}{2\overline{V}_D}\left[1+ 
    g_0^2\triangle^{(2)}+g_0^4\triangle^{(3)}+\dots\right]\,,
\end{equation}
(here $\overline{V}_D=\Ls^{d-1}\Lhat^q=\Ls^{D-1}\ellhat^q$), with
\begin{align}
  \triangle^{(2)}&=(n-2)R(0)\,,
  \\
  \triangle^{(3)}&=(n-2)\left[
    2W+\frac{3}{4\overline{V}_D}\IDR_{10:D-1}+(n-3)R(0)^2\right]
  -4\left(2l_1+nl_2\right)\ddot{R}(0)\,.
\end{align}
Here $R(z)$ is the propagator for an infinitely long strip without the slow
modes $\mathbf{p}=0$ \footnote{Our $R(z)$ is closely related to $\bar{G}^*(z)$
  of \cite{Has09}.}:
\begin{equation}
  R(z)=\frac{1}{2\overline{V}_D}\sum_{\bfp\ne0}
  \frac{\mre^{-\omega(\bfp)|z_0|+i\bfp\bfz}}{\omega(\bfp)}\,,
  \,\,\omega(\bfp)=\sqrt{\bfp^2}\,.
\end{equation}
The singularity of $R(z)$ at $z=0$ is regularized with DR.  Further
$\IDR_{10:D-1}$ is the regularized sum $\IDR_{10}$ in $D-1$ dimensions and
\begin{equation}
  W= -\int_{-\infty}^\infty\mrd z_0\,
  \int_{\mathbf{z}} R(z)^2\partial_0^2 R(z)\,.
\end{equation}
The computation of $W$ is the most involved part and we discuss this in detail
in \cite{Nie15b}.

Returning again to the case $d=4$, the moment of inertia has an expansion
\begin{equation}
  \frac{\Theta}{F^2\Ls^3}=1+\Theta_1^{\mathrm{DR}}(F\Ls)^{-2}
  +\Theta_2^{\mathrm{DR}}(F\Ls)^{-4}+\dots
\end{equation}
with
\begin{align}
  \Theta_1^{\mathrm{DR}}&=-(n-2)\Ls^2R(0)\,,
  \\
  \Theta_2^{\mathrm{DR}}&=(n-2)\Ls^4\left[
    -2W+R(0)^2-\frac{3}{4\overline{V}_D^2}
    \sum_{\bfp\ne0}\frac{1}{\bfp^2}\right]
  +4\left(2l_1+nl_2\right) \Ls^4 \ddot{R}(0)\,.
\end{align}

In \cite{Nie15b} we find
\begin{equation}
  W = \frac{5}{24\pi^2}\ddot{R}(0)\left[\frac{1}{D-4}-\ln L_s\right]
  + c_w\Ls^{-4}\,,
\end{equation}
with\footnote{The value $c_w=0.029492025146$ given in \cite{Has09} differs
  from ours.}
\begin{equation}
  c_w=0.0986829798\,.
\end{equation}
Adding the counter-terms using \eqref{liGL}, the $1/(D-4)$ singularities
cancel (and also the $\ellhat$-dependent terms coming from $\order{D-4}$
contributions in $\ddot{R}(0)$), and we obtain
\begin{equation}
  \Theta_2^{\mathrm{DR}}=(n-2)\theta_2
  +\frac{1}{12\pi^2}\Ls^4\ddot{R}(0)\left[
    (3n-10)\ln(\overline{c}\Lambda_1\Ls)
    +2n\ln(\overline{c}\Lambda_2\Ls)\right]\,,
\end{equation}
with\footnote{In ref.~\cite{Has09} the term in \eqref{Theta2final} involving
  $\beta_1^{(3)}$ is missing.}
\begin{equation}
  \theta_2=-2c_w+\Ls^4R(0)^2+\frac34\Ls\beta_1^{(3)}\,,
  \label{Theta2final}
\end{equation}
where in the notation of \cite{Has90}
\begin{equation}
  \beta_1^{(3)}=-\frac{1}{\overline{V}_D}\sum_{\bfp\ne0}\frac{1}{\bfp^2}\,,
\end{equation}
where the sum is over $3+q$ dimensional momenta $\bfp$.  With dimensional
regularization
\begin{equation}
  \beta_1^{(3)}=-\Ls R(0)\,.
\end{equation}

Putting in the numerical values \cite{Nie15b}
\begin{align}
  \Ls^2 R(0)&=-0.2257849594407580334832664917\,,
  \\
  \Ls^4\ddot{R}(0)&=-0.8375369106960818783868948293\,,
\end{align}
we obtain
\begin{equation}
  \theta_2=0.0229516079\,,
\end{equation}
completely consistent with the lattice result converted to DR in
\eqref{R2mLATT_DR1}. Note however that values of
$\theta_2,\overline{\theta}_2$ differ in the 6'th decimal place, which
indicates that at some stage(s) we have overestimated our numerical precision.

\section{Conclusions}
\label{Conclusions}

We have established relations between the 4-derivative couplings of effective
Lagrangians involving fields in the vector representation of O$(n)$ using both
lattice and dimensional regularizations.  This allows translation of results
obtained on the lattice to those of DR more commonly used in phenomenology.
Computations on the lattice are usually algebraically more complicated but
conceptually clear.

One application is to the computation of the mass gap of massless 2-flavor QCD
in the $\delta$--regime.  It is given by
\begin{equation}
  \begin{aligned}
    \Theta&=F^2\Ls^3\left[1+0.4515699182\,\frac{1}{F^2\Ls^2} \right.
    \\
    &\quad \left. +\frac{1}{F^4\Ls^4}\left(\theta-0.8375369109
        \frac{1}{6\pi^2}\left\{\ln(\Lambda_1\Ls)+4\ln(\Lambda_2\Ls)\right\}
      \right)+\ldots\right]\,,
  \end{aligned}
  \label{ThetaO4}
\end{equation}
with
\begin{equation}
  \begin{aligned}
    \theta&=2\theta_2+\frac{5}{6\pi^2}\Ls^4\ddot{R}(0)\overline{C}
    \\
    &=0.1503452489\,.
  \end{aligned}
\end{equation}
Note Hasenfratz \cite{Has09} obtained $\theta=0.088431628$.

It is convenient to rewrite \eqref{ThetaO4} by using the low-energy parameters
defined in \cite{Gas84},
\begin{equation}
  \overline{l}_i\equiv \ln \frac{\Lambda_i^2}{m_{\pi}^2}\,,
\end{equation}
where $m_\pi$ is the physical pion mass.  We have
\begin{equation}
  \begin{aligned}
    \Theta&=F^2\Ls^3\Bigg[ 1+\frac{0.45157}{F^2\Ls^2}
    \\
    &\quad +\frac{1}{F^4\Ls^4}\left( 0.1503-0.0283 \left[
        \overline{l}_2+\frac14 \overline{l}_1 +\frac52 \ln \left(L_s
          m_{\pi}\right)\right] \right) + \ldots \Bigg] \,.
  \end{aligned}
  \label{ThetaO4a}
\end{equation}

The QCDSF collaboration \cite{QCDSF11,Bie10,Bie11} compared their data for the
mass gap from numerical simulations of lattice QCD to \eqref{ThetaO4} using
values
\begin{equation}
  \overline{l}_1=-0.4\pm 0.6\,, \qquad 
  \overline{l}_2=4.3 \pm 0.1\,. 
\end{equation}
taken from \cite{Col01}.  They found satisfactory agreement with the analytic
result and our new value for $\theta$ doesn't change this conclusion.

Although measuring the low lying spectrum is among the simplest and cleanest
numerical problems, a difficulty is that the box size needs to reach $3\,\fm$
or larger.  This is suggested by the NLO 
correction which is $38\%,\,26\%$ and
$15\%$ of the leading order for $L_s=2.5,\,3,\,4$ fermi respectively, where
for the estimates we have used the value $F=86.2$ MeV from Colangelo and
D\"{u}rr \cite{Col03}.  The NNLO correction is unexpectedly small:
$-0.6\%,\,-0.7\%$ and $-0.5\%$ at the same lattice sizes.  Note however, that
this is due to the cancellation of the two terms in \eqref{ThetaO4a}, and the
smallness of the NNLO correction does not indicate the smallness 
of the next, unknown correction.

Note that the combination $\overline{l}_2+\overline{l}_1/4$ enters with a
small coefficient, whose value e.g.\ for $L_s=3\,\mathrm{fm}$ is $-0.0095$. As
a consequence, the mass gap is not sensitive to these parameters. For the same
reason, however, it provides a clean way to obtain the value of $F$,
in particular the constant $\overline{l}_4$ which controls the ratio
$F_\pi/F$ close to the chiral limit.
At the physical pion mass the sensitivity of the mass gap in the
delta regime to this parameter is roughly $0.2\, \overline{l}_4$.
Alternatively, knowing $F$, one can estimate the corresponding 
lattice artifacts, the goodness of the chiral extrapolation, etc.

Numerical simulations of lattice QCD in the $\delta$--regime potentially still
give a good possibility to constrain the LE constants of $\chi PT$. The mass
gap is unfortunately only sensitive to the decay constant $F$, and it remains
a challenge to find other observables which are sensitive to the $\bar{l}_i$
and also accurately measurable in numerical simulations.

The susceptibility discussed in this paper could also serve as
such an alternative quantity. To use this one has to consider massless 
Ginsparg-Wilson fermions (having exact chiral symmetry 
\cite{Luscher:1998pqa})
and using the associated conserved currents 
\cite{Kikukawa:1998py,Hasenfratz:2002rp}.
This choice has the advantage that one has an extra parameter, $\ell$,
which affects the sensitivity to the LEC's $\overline{l}_1$ and
$\overline{l}_2$.
Writing the susceptibility for $n=4$ as
\begin{equation}
  \chi = \frac{1}{2 F^2} \left[1 + \frac{1}{F^2 L^2} e_1
    +   \frac{1}{F^4 L^4} \left(e_2 + d_1 \overline{l}_1
      +  d_2 \overline{l}_2\right) + \order{\frac{1}{F^6 L^6}} \right]\,, 
  \label{chi_d4}
\end{equation}
one has (cf. \eqref{chiDR}, \eqref{R1f} and \eqref{R2tot})
\begin{equation}
  \begin{aligned}
    e_1 & = -\frac{1}{2\pi}(\gamma_2 - 1) \,, 
    \\
    d_1 & = -\frac{1}{24\pi^2}
    \left(\gamma_1 - \frac{3}{\ell} - \frac12\right)\,, 
    \\
    d_2 & = -\frac{1}{6\pi^2}
    \left( \gamma_1 - \frac{1}{2\ell} - \frac12\right) \,.
  \end{aligned}
  \label{edd}
\end{equation}
The shape dependence of the coefficients $e_1$, $d_1$, $d_2$ is 
illustrated\footnote{We did not calculate the 2-loop contribution 
$\mcWov(\ell)$ (needed for $e_2$) for $\ell < 1$.}
in table~\ref{e1d1d2}.
As one can see, the sensitivity to $\overline{l}_i$ increases with 
decreasing $\ell$. This can perhaps be understood: at $h>0$ for large $\ell$
(large time direction)
the state with maximal $Q_{12}$ in an isospin multiplet is dominating,
while decreasing $\ell$ all other states from the given multiplets 
start to influence the free energy, and because of nearly degenerate 
states this increases the susceptibility. 
Note, however, that the NLO coefficient $e_1(\ell)$ also increases
(although not as fast as the NNLO coefficients)
and at fixed size $L_s=L$ one cannot trust the expansion for too small
values of $\ell$.

\begin{table}[ht]
  \centering
  \begin{tabular}[t]{|l|c|c|c|}
    \hline
    $\ell$ & $e_1$ & $d_1$ & $d_2$ \\[1.0ex]
    \hline 
    3    &  $-0.548430$      & $    -0.000035$    & $-0.014214$ \\
    2    &  $-0.215019$      & $\phm 0.003484$    & $-0.007174$ \\
    1    &  $\phm 0.140461$  & $\phm 0.014776$    & $\phm 0.016887$  \\
    0.5  &  $\phm 0.666497$  & $\phm 0.069766$    & $\phm 0.194629$   \\  
    0.25 &  $\phm 2.666667$  & $\phm 0.761772$    & $\phm 2.878218$    \\
    \hline
  \end{tabular}
  \caption{\footnotesize The shape dependence of the coefficients
    in \eqref{edd}.} 
  \label{e1d1d2}
\end{table}

\section*{Acknowledgments}

Christoph Weiermann participated at an early stage of these calculations.  We
thank him for the collaboration.  We also would like to thank Janos Balog,
Gilberto Colangelo, Peter Hasenfratz, Heiri Leutwyler
and Martin L\"uscher for useful discussions.

\begin{appendix}

  \section{\boldmath The terms $B_4^{(i)}$ and $C_4^{(i)}$}
  \label{AppA}

  The terms $B_4^{(i)}$ and $C_4^{(i)}$ appearing in (\ref{A4h}) are given by
  \begin{align}
    B_4^{(1)} & =B_4^{(5a)}+B_4^{(1a)}\,,
    \\
    B_4^{(1a)} & =4\sum_{x k}(\Box_k S_1\nabla_0 S_2-\Box_k S_2\nabla_0
    S_1)\,,
    \\
    C_4^{(1)} & =C_4^{(5a)}+C_4^{(1a)}\,,
    \\
    C_4^{(1a)} & =\sum_{x k}\left\{(2S_1+\Box_0 S_1)\Box_k S_1 +(2S_2+\Box_0
      S_2)\Box_k S_2\right\}\,,
  \end{align}
  where $\nabla_0 = \frac12 (\pz + \pz^*)$ is the symmetric derivative and
  $\Box_\mu = \partial_\mu \partial_\mu^\star$. 
  Here and in the rest of this section we
  have suppressed the argument of the fields e.g. $S_i=S_i(x)$. Also below we
  introduce the notation $S'_i=S_i(x+\hat{0})$ and $S''_i=S_i(x+2\hat{0})$.
  \begin{align}
    B_4^{(2)} & =-4\sum_x (S_1 S'_2-S_2 S'_1)\pmu \bS \cdot \pmu \bS\,,
    \\
    C_4^{(2)} & =-\sum_x\left\{4(S_1 S_2'-S_2 S_1')^2 +2(S_1 S_1' + S_2 S_2')
      \pmu \bS\cdot \pmu \bS\right\}\,.
  \end{align}

\begin{align}
  B_4^{(3)} & =B_4^{(4a)}+B_4^{(3a)}\,,
  \\
  B_4^{(3a)} & =-4\sum_{x k}(S'_1 \partial_k S_2 - S'_2 \partial_k S_1) \pz\bS
  \cdot \partial_k \bS\,,
  \\
  C_4^{(3)} & =C_4^{(4a)}+C_4^{(3a)}\,,
  \\
  C_4^{(3a)} & =2\sum_{x k}\left[(S'_1 \partial_k S_1 + S'_2 \partial_k S_2)
    \pz \bS \cdot \partial_k \bS -(S'_1 \partial_k S_2-S'_2\partial_k
    S_1)^2\right]\,.
\end{align}

\begin{align}
  B_4^{(4a)} & =-4\sum_x(S_1 S_2'-S_2 S_1')\pz\bS \cdot \pz\bS\,,
  \\
  C_4^{(4a)} & =-\sum_x\left\{2(S_1 S_1' + S_2 S_2')\pz\bS \cdot \pz\bS
    +4(S_1S_2'-S_2 S_1')^2\right\}\,.
\end{align}

\begin{align}
  B_4^{(5a)} & =-4\sum_x\left( \pz^* S_1 \pz S_2 - \pz^* S_2 \pz S_1\right)
  \,,
  \\
  C_4^{(5a)} & =-2\sum_x\left\{ -(S_1 \Box_0 S_1 + S_2 \Box_0 S_2) + 2(\pz S_1
    \pz^* S_1 + \pz S_2 \pz^* S_2 )\right\}\,,
\end{align}

\begin{align}
  B_4^{(5c)} & =B_4^{(5a)}+B_4^{(5d)}\,,
  \\
  B_4^{(5d)} & = -4\sum_x\left\{[S_1''-S_1']\partial_k^2S_2
    -[S_2''-S_2']\partial_k^2S_1\right\}\,,
  \\
  C_4^{(5c)} & =C_4^{(5a)}+C_4^{(5d)}\,,
  \\
  C_4^{(5d)} & = 2\sum_{x k}\left\{[2S_1''-S_1']\partial_k^2S_1
    +[2S_2''-S_2']\partial_k^2S_2\right\}\,.
\end{align}

\section{Some lattice momentum sums}
\label{AppB}

We define the following lattice sums:
\begin{align}
  I_{nm} & \equiv \frac{1}{V} \psump
  \frac{\left(\hat{p}_0^2\right)^m}{\left(\hat{p}^2\right)^n}\,, \label{Inm}
  \\
  J_{nm} & \equiv \frac{1}{V}\psump\frac{(\hat{p}_0^2)^m
    \sum_\mu\hat{p}_\mu^4}{(\hat{p}^2)^n}\,, \label{Jnm}
  \\
  K_{nm} & \equiv \frac{1}{V}\psump\frac{(\hat{p}_0^2)^m
    (\sum_\mu\hat{p}_\mu^4)^2}{(\hat{p}^2)^n}\,, \label{Knm}
  \\
  L_{nm} & \equiv \frac{1}{V}\psump\frac{(\hat{p}_0^2)^m}{(\hat{p}^2)^n}
  \sum_{\mu\nu} \cos(p_\mu-p_\nu) \phat_\mu^2 \phat_\nu^2 \,, \label{Lnm}
  \\
  J_{nmk} & \equiv \frac{1}{V}\psump\frac{(\hat{p}_0^2)^m
    \sum_\mu\hat{p}_\mu^{2k}}{(\hat{p}^2)^n}\,.  \label{Jnmk}
\end{align}

The following momentum sums which appear in our computation are expressed in
terms of these:
\begin{equation}
  \begin{split}
    \mcF_1 & \equiv \frac{1}{V^2} \psum{p q}
    \frac{\hat{p}_0^2 (\widehat{p+q})^2}{(\hat{p}^2)^2 \hat{q}^2 } 
    \\
    & = I_{11} I_{10} + I_{21} I_{00} -\frac{\ds+1}{2\ds} I_{22} I_{11} -
    \frac{1}{2\ds} I_{11} I_{00} +\frac{1}{2\ds} I_{11}^2 + \frac{1}{2\ds}
    I_{22} I_{00}\,.
  \end{split}
  \label{cF1}
\end{equation}

\begin{equation}
  \begin{split}
    \mcF_2 & \equiv \frac{1}{V^2} \psum{p q} \frac{\sin^2 p_0
      \left(\widehat{p+q} \right)^2}{\left(\hat{p}^2\right)^3\hat{q}^2 }
    \\
    & = I_{10}\left( I_{21}-\frac14 I_{22} \right) + I_{00} \left(
      I_{31}-\frac14 I_{32} \right) -\frac12 I_{11} \left( I_{32}-\frac14
      I_{33} \right)
    \\
    & \qquad +\frac{1}{2 d_s} \left( I_{11}-I_{00}\right) \left(
      I_{21}-\frac14 I_{22} - I_{32} + \frac14 I_{33} \right) \,.
  \end{split}
  \label{cF2}
\end{equation}

The expressions for $\mcF_1$ and $\mcF_2$ are valid for a spatially symmetric
volume, $V=\Ls^{\ds} \Lt$.

\begin{equation}
  \begin{split}
    \mcF_3 & \equiv \frac{1}{V^2} \psum{p q} \frac{\sin p_0 \sin q_0
      \left(\widehat{p+q} \right)^2}{\left(\hat{p}^2\right)^2
      \left(\hat{q}^2\right)^2 }  \\
    & = 2 \left( \frac{1}{V} \psump \frac{\sin^2
        p_0}{\left(\hat{p}^2\right)^2}\right)^2 = 2 \left( I_{21} -\frac14
      I_{22} \right)^2\,.
  \end{split}
  \label{cF3}
\end{equation}

\begin{equation}
  \begin{split}
    \mcF_4 & = L_{21} = \frac{1}{V}\psump \frac{\hat{p}_0^2}{(\hat{p}^2)^2}
    \sum_{\mu\nu}\cos(p_\mu-p_\nu) \hat{p}_\mu^2\hat{p}_\nu^2 \\
    & = I_{01}-J_{11}+\frac14 K_{21}+ J_{213}-\frac14 J_{214}\,.
  \end{split}
  \label{cF4}
\end{equation}

\begin{equation}
  \begin{split}
    \mcF_5 & = \frac{1}{V}\psump \frac{\sin^2p_0}{(\hat{p}^2)^3} \sum_{\mu\nu}
    \cos(p_\mu-p_\nu)\hat{p}_\mu^2\hat{p}_\nu^2 = L_{31}-\frac14 L_{32}
    \\
    & = I_{11}-J_{21}+\frac14 K_{31} + J_{313} - \frac14 J_{314}
    -\frac14\left[ I_{12}-J_{22}+\frac14 K_{32} + J_{323} - \frac14 J_{324}
    \right]\,.
  \end{split}
  \label{cF5}
\end{equation}

\begin{equation}
  \begin{split}
    \mcF_6 & \equiv \frac{1}{V}\psump\frac{1}{\hat{p}^2}
    \sum_k\left[2\cos(2p_0)-\cos(p_0)\right]\cos(p_k)\hat{p}_k^2
    \\
    & = I_{00}-\frac72 I_{01}+I_{02}-\frac12 J_{10}+\frac74 J_{11} -\frac12
    J_{12} -I_{11}+4I_{12}-\frac{11}{4}I_{13}+\frac12 I_{14}\,.
  \end{split}
  \label{cF6}
\end{equation}

\begin{equation}
  \begin{split}
    \mcF_7 & \equiv \frac{1}{V}\psump\sum_k \frac{\left(\mre^{-ip_0}-1\right)
      \left(\mre^{-i2p_0}-1\right)\left(\mre^{ip_k}-1\right)^2}{(\hat{p}^2)^2}
    \\
    & = -\frac{1}{V}\psump\sum_k \frac{\hat{p}_k^2\cos p_k \left(\cos
        3p_0-\cos 2p_0-\cos p_0+1\right)}{(\hat{p}^2)^2}
    \\
    & = 2I_{11}-\frac52 I_{12}+\frac12 I_{13}-2 I_{22} +\frac72 I_{23}-\frac74
    I_{24}+\frac14 I_{25} -J_{21}+\frac54 J_{22}-\frac14 J_{23}\,.
  \end{split}
  \label{cF7}
\end{equation}

\section{\boldmath Correlators appearing in $F_r$ with lattice regularization}
\label{AppC}

\subsection{\boldmath Correlators appearing in $F_{1,0}$ and $F_{1,1}$}
\label{AppC1}

First
\begin{equation}
  \langle U_1 \rangle_0 = n_1 \frac{1}{V} \psump
  \frac{\hat{p}_0^2}{\hat{p}^2} = n_1 I_{11}\,,
  \label{U1av} 
\end{equation}
where $I_{nm}$ are defined in \eqref{Inm}. Next
\begin{equation}
  \begin{aligned}
    \langle U_2 \rangle_0 &= 
    \pz^x \pz^y \left. \frac{1}{4V} \sum_x  
      \langle \vpix^2 \vpiy^2 \rangle_0 \right|_{y=x} = 
    \frac{n_1}{2}\pz^x \pz^y \left. \Bigl[
      \frac{1}{V} \sum_x G(x-y)^2 \Bigr] \right|_{y=x}   \label{U20}
    \\
    &= \frac{n_1}{V^2}\psum{p q}\frac{1-\cos(p_0+q_0)}{\hat{p}^2\hat{q}^2} 
    = n_1 I_{11}\left( I_{10} - \frac14 I_{11}\right)\,.  
  \end{aligned}
\end{equation}

\begin{align}
  \langle U_1 A_{2,0}\rangle_0^c & = n_1 I_{11} \,, \label{U1A0}
  \\
  \langle U_1 A_{2,1}^{(a)}\rangle_0^c &= - n_1 \left( 1-\frac{n_1}{V}\right)
  I_{21}\,, \label{U1A1a}
\end{align}
\begin{equation}
  \begin{aligned}
    \langle U_1 A_{2,1}^{(b)} \rangle_0^c &= \left.\frac{1}{8V}\sum_{x
        u}\pmu^u\pmu^v \langle \pz\vpix \cdot \pz \vpix
      \vpiu^2\vpiv^2\rangle_0^c\right|_{v=u}
    \\
    &=\left.\frac{n_1}{V}\sum_{x u}\pmu^u\pmu^v\left\{
        \pz^xG(x-u)\pz^xG(x-v)G(u-v)\right\}\right|_{v=u}
    \\
    &= n_1 \mcF_1\,,
  \end{aligned}
  \label{U1A1b}
\end{equation}
where $\mcF_1$ is given by \eqref{cF1}. Also\footnote{Note that one can
  obtain the results of insertions in eqs.~\eqref{U1A0} and
  \eqref{U1A41}-\eqref{U1A415c} by observing that $\langle X
  A_{2,0}\rangle_0^c$ inserts for each propagator appearing in $\langle X
  \rangle_0$ a factor 1, i.e.\ simply counts the number of propagators in
  $\langle X \rangle_0$.  Similarly, for the other operators the corresponding
  insertions are $A_{4,1}^{(1)} \to 2\phat^2$, $A_{4,1}^{(5a)} \to 2\sum_\mu
  \phat_\mu^4 / \phat^2$, and $A_{4,1}^{(5c)} \to
  2\sum_{\mu\nu}\cos(p_\mu-p_\nu) \phat_\mu^2\phat_\nu^2/ \phat^2$.}
\begin{align}
  \langle U_1A_{4,1}^{(1)}\rangle_0^c & =2n_1 I_{01}\,, \label{U1A41}
  \\
  \langle U_1A_{4,1}^{(i)}\rangle_0^c & =0\,,\,\,\,\,i=2,3,4\,, \label{U1A41i}
  \\
  \langle U_1A_{4,1}^{(5a)}\rangle_0^c & =2n_1J_{21}\,, \label{U1A415a}
  \\
  \langle U_1A_{4,1}^{(5b)}\rangle_0^c & =2n_1 I_{01}\,, \label{U1A415b}
  \\
  \langle U_1A_{4,1}^{(5c)}\rangle_0^c & =2n_1\sum_x \pz^*\pmu\pmu
  G(x)\pz^*\pnu\pnu G(x) =2n_1\mcF_4\,, \label{U1A415c}
\end{align}
where $\mcF_4$ is given by \eqref{cF4}.

\subsection{\boldmath Correlators appearing in $F_{2,0}$ and $F_{2,1}$}
\label{AppC2}

Firstly
\begin{equation}
  \langle W_2 \rangle_0 = n_1(n_1-1) \left( I_{21}
    - \frac14 I_{22}\right)\,.
  \label{W2av}
\end{equation}
Next
\begin{equation}
  \begin{aligned}
    \langle W_3\rangle_0 &= \frac{n_1}{V}\sum_{x y}G(x-y) 
    \left\{ n_1 G(x-y)\nabla_0^x\nabla_0^y G(x-y) 
      +  2\left[\nabla_0^x G(x-y)\right]
      \left[\nabla_0^y G(x-y)\right]\right\}   \label{W3avC}
    \\
    &= -n_1^2 \sum_x G(x)^2\nabla_0^2 G(x) 
    - 2 n_1 \sum_x \left[\nabla_0 G(x)\right]^2 G(x)  
    \\
    &= n_1(n_1-1)  W_{3a} + \frac12 n_1  W_{3c}\,, 
  \end{aligned}
\end{equation}
where $W_{3a},W_{3c}$ are defined in \eqref{W3a},\eqref{W3c} respectively. For
the connected correlators we get
\begin{equation}
  \begin{aligned}
    \langle W_2 A_{2,0} \rangle_0^c &= n_1\sum_{x y
      \mu}\Bigl\{n_1G(x-y)\pmu^*\nabla_0G(x-z) \pmu^*\nabla_0G(y-z)
    \\
    & \quad -n_1\nabla_0\nabla_0G(x-y)\pmu^*G(x-z)\pmu^*G(y-z)
    \\
    &\quad +2\nabla_0G(x-y)\pmu^*\nabla_0G(x-z)\pmu^*G(y-z)\Bigr\}
    \\
    &=2n_1(n_1-1)\left[ I_{21}-\frac14 I_{22}\right]\,.
  \end{aligned}
  \label{W2A0}
\end{equation}
\begin{equation}
  \begin{aligned} 
    \langle W_2 A_{2,1}^{(a)}\rangle_0^c &= -2n_1(n_1-1) \left(1 -
      \frac{n_1}{V}\right) \frac{1}{V}\sum_{x y u}
    G(x-u)G(y-u)\nabla_0^x\nabla_0^y G(x-y)
    \\
    &= -2n_1(n_1-1) \left(1 - \frac{n_1}{V} \right) \frac{1}{V} \psump
    \frac{\sin^2 p_0}{\left(\hat{p}^2\right)^3}
    \\
    &= - 2n_1(n_1-1) \left(1 - \frac{n_1}{V} \right) \left(I_{31} -\frac14
      I_{32} \right)\,.
  \end{aligned}
  \label{W2A1aaveval}
\end{equation}
\begin{equation}
  \begin{aligned}
    \langle W_2 A_{2,1}^{(b)} \rangle_0^c &= \frac{1}{8V}\sum_{x y
      u}\pmu^u\pmu^v \left.\left\langle \left[\nabla_0\vpix\cdot
          \nabla_0\vpiy\right]\vpix\cdot\vpiy
        \vpiu^2\vpiv^2\right\rangle_0^c\right|_{v=u}  \\
    &= n_1(n_1-1)\frac{1}{V}\sum_{x y u} \pmu^u\pmu^v\Bigl[ G(x-u)G(y-v)\times
    \\
    &\left\{ 2G(u-v)\nabla_0^x\nabla_0^yG(x-y)
      -\nabla_0^xG(x-v)\nabla_0^yG(y-u)\right\}
    \Bigr]_{v=u}  \\
    & = n_1(n_1-1)\left[ 2\mcF_2- \mcF_3\right]\,,
  \end{aligned}
  \label{W2A1baveval}
\end{equation}
with $\mcF_2,\mcF_3$ defined in \eqref{cF2},\eqref{cF3}.

\begin{align}
  \langle W_2A_{4,1}^{(1)}\rangle_0^c & =4n_1(n_1-1)\left[I_{11}-\frac14
    I_{12}\right]\,, \label{W2A41}
  \\
  \langle W_2A_{4,1}^{(i)}\rangle_0^c & =0\,,\,\,\,\,i=2,3,4\,, \label{W2A4i}
  \\
  \langle W_2A_{4,1}^{(5a)}\rangle_0^c & =4n_1(n_1-1)\left[J_{31}-\frac14
    J_{32}\right]\,, \label{W2A45a}
  \\
  \langle W_2A_{4,1}^{(5b)}\rangle_0^c & =4n_1(n_1-1)\left[I_{11}-\frac14
    I_{12}\right]\,,\label{W2A45b}
  \\
  \langle W_2A_{4,1}^{(5c)}\rangle_0^c & =4n_1(n_1-1)\mcF_5 \,, \label{W2A45c}
\end{align}
where $\mcF_5$ is given by \eqref{cF5}.

\subsection{\boldmath Computation of $F_3$ up to $\order{g_0^2}$}
\label{AppC3}

We have
\begin{equation}
  F_3 =\sum_{i=1}^5\frac{g_4^{(i)}}{4}F_3^{(i)}\,,
  \,\,\,\,\,\,\,\,
  F_3^{(i)} = \frac{1}{V}\langle C_4^{(i)}\rangle_\mcA\,.
\end{equation}
Averaging over the rotations gives the following expressions for the
$F_3^{(i)}$:
\begin{align}
  F_3^{(1)} & =F_3^{(5a)}+F_3^{(1a)}\,,
  \\
  F_3^{(1a)} & =\frac{2}{n}\frac{1}{V} \sum_{x
    k}\langle\left[2\bSx+\Box_0\bSx\right]\cdot\Box_k\bSx \rangle_\mcA\,.
\end{align}
\begin{equation}
  F_3^{(2)}=-\frac{4}{nn_1}\frac{1}{V}\sum_x
  \langle 2-2\{\bSx\cdot\bS'(x)\}^2
  +n_1\{\bSx\cdot\bS'(x)\}\pmu \bSx\cdot \pmu\bSx
  \rangle_\mcA\,,  
\end{equation}
where we have introduced the notation $S'_i(x)=S_i(x+\hat{0})$ and below
$S''_i(x)=S_i(x+2\hat{0})$.
\begin{align}
  F_3^{(3)} & =F_3^{(4a)}+F_3^{(3a)}\,,
  \\
  F_3^{(3a)} & =\frac{4}{nn_1}\frac{1}{V} \sum_{xk}\left\langle
    \bS'(x)\cdot\partial_k\bSx
    \left[n_1\pz\bSx\cdot\partial_k\bSx+\bS'(x)\cdot\partial_k\bSx\right]
  \right. \\
  & \qquad \left.
    -\partial_k\bSx\cdot\partial_k\bSx\right\rangle_\mcA\,. \nonumber
\end{align}

\begin{equation}
  F_3^{(4)}= F_3^{(4a)}-\frac{1}{d+2}\left(F_3^{(2)} +2F_3^{(3)}\right)\,,
\end{equation}
with
\begin{equation}
  F_3^{(4a)}=-\frac{4}{nn_1}\frac{1}{V}
  \sum_x\langle\left\{n_1 (\bSx\cdot\bS'(x)) \pz\bSx\cdot\pz\bSx+2
    -2(\bSx\cdot\bS'(x))^2\right\}
  \rangle_\mcA\,. 
\end{equation}
Next
\begin{equation}
  F_3^{(5a)}=-\frac{4}{n}\frac{1}{V}\sum_x\langle
  -\bSx\cdot\Box_0\bSx+2\pz\bSx\cdot\pz^*\bSx\rangle_\mcA\,. 
\end{equation}
\begin{equation}
  F_3^{(5b)}=F_3^{(1)}\,. 
\end{equation}
\begin{align}
  F_3^{(5c)} & =F_3^{(5a)}+F_3^{(5d)}\,,
  \\
  F_3^{(5d)} & =\frac{4}{n}\frac{1}{V}\sum_{x k}\langle
  \left[2\bS''(x)-\bS'(x)\right]\cdot
  \partial_k^2\bSx\rangle_\mcA\,.
\end{align}

$F_3$ has a perturbative expansion starting at $\order{g_0^2}$:
\begin{equation}
  F_3  =\sum_{r=1}F_{3,r}g_0^{2r}\,,\,\,\,\,\,\,
  F_3^{(i)}=\sum_{r=1}F_{3,r}^{(i)}g_0^{2r}\,.
\end{equation}

\subsubsection{\boldmath Computation of $F_{3,1}^{(i)}$}

\begin{align}
  F_{3,1}^{(1)} & =F_{3,1}^{(5a)}+F_{3,1}^{(1a)}\,,
  \\
  F_{3,1}^{(1a)} & =\frac{2}{n}\frac{1}{V} \sum_{x
    k}\langle\left[2\vpix+\Box_0\vpix\right]\cdot\Box_k\vpix\rangle_0
  \nonumber \\
  & =\frac{2n_1}{n}\sum_k\left[2+\Box_0\right]\Box_kG(0)
  \nonumber \\
  & =\frac{2n_1}{n}\left[-2I_{00}+2I_{11}+I_{01}-I_{12}\right]\,.
\end{align}
\begin{equation}
  \begin{aligned}
    F_{3,1}^{(2)} & =-\frac{4}{nn_1}\frac{1}{V}\sum_x \langle
    2\pz\vpix\cdot\pz\vpix+n_1\pmu\vpix\cdot\pmu\vpix\rangle_0
    \\
    & =\frac{4}{n}\left\{2\Box_0 G(0)+n_1\Box G(0)\right\}
    \\
    & =-\frac{4}{n}\left\{2I_{11}+n_1I_{00}\right\}\,.
  \end{aligned}
\end{equation}
\begin{align}
  F_{3,1}^{(3)} & =F_{3,1}^{(4a)}+F_{3,1}^{(3a)}\,,
  \\
  F_{3,1}^{(3a)} & =-\frac{4}{nn_1}\frac{1}{V} \sum_{x
    k}\langle\partial_k\vpix\cdot\partial_k\vpix\rangle_0
  \nonumber \\
  & =-\frac{4}{n}\left[I_{00}-I_{11}\right]\,.
\end{align}
\begin{equation}
  \begin{aligned}
    F_{3,1}^{(4a)} & =-\frac{4(n_1+2)}{nn_1}\frac{1}{V}
    \sum_x\langle\pz\vpix\cdot\pz\vpix\rangle_0
    \\
    & =-\frac{4(n_1+2)}{n}I_{11}\,.
  \end{aligned}
\end{equation}
\begin{equation}
  \begin{aligned}
    F_{3,1}^{(5a)} & =-\frac{4}{n}\frac{1}{V}\sum_x\langle
    -\vpix\cdot\Box_0\vpix+2\pz\vpix\cdot\pz^*\vpix\rangle_0
    \\
    & =\frac{4n_1}{n}\left\{\Box_0 G(0)+2\pz^2G(0)\right\}
    \\
    & =-\frac{4n_1}{n}\left\{3I_{11}-I_{12}\right\}\,.
  \end{aligned}
\end{equation}
Finally
\begin{align}
  F_{3,1}^{(5c)} & =F_{3,1}^{(5a)}+F_{3,1}^{(5d)}\,,
  \\
  F_{3,1}^{(5d)} & =\frac{4}{n}\frac{1}{V} \sum_{x
    k}\langle\left[2\vpi''(x)-\vpi'(x)\right] \cdot\partial_k^2\vpix\rangle_0
  \nonumber \\
  & =-\frac{4n_1}{n}\mcF_6\,,
\end{align}
where $\mcF_6$ is given by \eqref{cF6}.

\subsection{\boldmath Computation of $F_4$ up to $\order{g_0^2}$}
\label{AppC4}

$F_4$ is given by
\begin{equation}
  F_4 =Z_4\sum_i\frac{g_4^{(i)}}{4}F_4^{(i)}\,,
  \,\,\,\,\,\,\,
  F_4^{(i)}  = \frac{1}{V}\langle BB_4^{(i)}\rangle_\mcA\,.
\end{equation}
Again averaging over rotations:
\begin{align}
  F_4^{(1)} & =F_4^{(5a)}+F_4^{(1a)}\,,
  \\
  F_4^{(1a)} & =-\frac{8}{nn_1g_0^2}\frac{1}{V} \sum_{x y k}\langle
  \left[\bSx\cdot\Box_k\bSy\right]\bS'(x)\cdot\nabla_0\bSy \nonumber
  \\ 
  & \qquad -\left[\bSx\cdot\nabla_0\bSy\right]\bS'(x)\cdot\Box_k\bSy \rangle_\mcA\,.
\end{align}
\begin{equation}
\begin{aligned}
  F_4^{(2)}&=\frac{8}{nn_1g_0^2}\frac{1}{V}\sum_{x y} \langle
  \Bigl[\{\bSx\cdot\bSy\}\bS'(x)\cdot\bS'(y)
  \\
  &\qquad -\{\bSx\cdot\bS'(y)\}\bS'(x)\cdot\bSy\Bigr] \pmu\bSy\cdot\pmu\bS(y)
  \rangle_\mcA\,.
\end{aligned}
\end{equation}
\begin{align}
  F_4^{(3)} & =F_4^{(4a)}+F_4^{(3a)}\,,
  \\
  F_4^{(3a)} & =\frac{8}{nn_1g_0^2}\frac{1}{V} \sum_{x y k} \langle\Bigl[
  \{\bSx\cdot\bS'(y)\}\bS'(x)\cdot\partial_k\bSy
  \nonumber\\
  &\quad -\{\bS'(x)\cdot\bS'(y)\}\bSx\cdot\partial_k\bSy\Bigr]
  \pz\bSy\cdot\partial_k\bSy \rangle_\mcA\,.
\end{align}
\begin{equation}
\begin{aligned}
  F_4^{(4a)}&=\frac{8}{nn_1g_0^2}\frac{1}{V} \sum_{x y}\langle \Bigl[
  \{\bSx\cdot\bSy\}\bS'(x)\cdot\bS'(y)
  \\
  & \qquad -\{\bSx\cdot\bS'(y)\}\bS'(x)\cdot\bSy\Bigr] \pz\bSy\cdot\pz\bSy
  \rangle_\mcA\,.
\end{aligned}
\end{equation}
\begin{equation}
  F_4^{(5a)}=\frac{8}{nn_1g_0^2}\frac{1}{V}\sum_{x y}\langle
  \{\bSx\cdot\pz^*\bSy\}\bS'(x)\cdot\pz\bSy
  -\{\bSx\cdot\pz\bSy\}\bS'(x)\cdot\pz^*\bSy
  \rangle_\mcA\,. 
\end{equation}
\begin{equation}
  F_4^{(5b)}=F_4^{(1)}\,. 
\end{equation}
Finally
\begin{align}
  F_4^{(5c)} & =F_4^{(5a)}+F_4^{(5d)}\,,
  \\
  F_4^{(5d)} & = \frac{8}{nn_1g_0^2}\frac{1}{V}\sum_{x y k}\langle
  \{\bSx\cdot\left[\bS''(y)-\bS'(y)\right]\}\bS'(x)\cdot\partial_k^2\bSy
  \nonumber\\
  &-\{\bS'(x)\cdot\left[\bS''(y)-\bS'(y)\right]\}\bSx\cdot\partial_k^2\bSy
  \rangle_\mcA\,.
\end{align}

$F_4$ has a perturbative expansion starting at $\order{g_0^2}$:
\begin{equation}
  F_4  =\sum_{r=1}F_{4,r}g_0^{2r}\,,\,\,\,\,\,\,
  F_4^{(i)}=\sum_{r=1}F_{4,r}^{(i)}g_0^{2r}\,.
\end{equation}

\subsubsection{\boldmath Computation of $F_{4,1}^{(i)}$}

\begin{align}
  F_{4,1}^{(1)} & =F_{4,1}^{(5a)}+F_{4,1}^{(1a)}\,,
  \\
  F_{4,1}^{(1a)} & =-\frac{4}{nn_1}\frac{1}{V} \sum_{x y
    k}\langle\Box_k^y\left[(\vpix-\vpiy)^2\right]
  \nabla_0^y\left[(\vpi'(x)-\vpiy)^2\right]\rangle_0
  \nonumber \\
  & =-\frac{8}{n}\frac{1}{V}\sum_{x y k}\Bigl[2n_1\Box_k^yG(x-y)
  \nabla_0^yG(x+\hat{0}-y)
  \nonumber \\
  &\quad +\left(\Box_k^y\nabla_0^z\left\{G(\hat{0})-G(x-z)
      -G(y-x-\hat{0})+G(y-z)\right\}^2 \right)_{z=y}\Bigr]
  \nonumber \\
  & =\frac{16(n_1-1)}{n} \left\{I_{11}-\frac14 I_{12}-I_{22}+\frac14
    I_{23}\right\}\,.
\end{align}
\begin{equation}
  F_{4,1}^{(2)}=\frac{8}{nn_1}\frac{1}{V}\sum_{x y}
  \langle\{\pz\vpix\cdot\pz\vpiy\}\pmu\vpiy\cdot\pmu\vpiy
  \rangle_0=0\,. 
\end{equation}
\begin{align}
  F_{4,1}^{(3)} & =F_{4,1}^{(4a)}+F_{4,1}^{(3a)}\,,
  \\
  F_{4,1}^{(3a)} & =\frac{8}{nn_1}\frac{1}{V} \sum_{x y k}\langle
  \{\pz\vpix\cdot\partial_k\vpiy\}\pz\vpiy\cdot\partial_k\vpiy \rangle_0=0\,.
\end{align}
\begin{equation}
  F_{4,1}^{(4a)}=\frac{8}{nn_1}\frac{1}{V}
  \sum_{x y}\langle
  \{\pz\vpix\cdot\pz\vpiy\}\pz\vpiy\cdot\pz\vpiy\rangle_0=0\,. 
\end{equation}
\begin{equation}
  \begin{aligned}
    F_{4,1}^{(5a)} & =\frac{8}{nn_1}\frac{1}{V}\sum_{x y}\left\langle
    \{\vpix\cdot\pz^*\vpiy\}\vpi'(x)\cdot\pz\vpiy
    -\{\vpix\cdot\pz\vpiy\}\vpi'(x)\cdot\pz^*\vpiy\right\rangle_0
    \\
    & =\frac{8(n_1-1)}{n}\frac{1}{V}\sum_{x y}
    \left\{\pz^{*y}G(x-y)\pz^yG(x+\hat{0}-y)
      -\pz^yG(x-y)\pz^{*y}G(x+\hat{0}-y)\right\}
    \\
    & =\frac{16(n_1-1)}{n}\left\{I_{22}-\frac14 I_{23}\right\}\,,
  \end{aligned}
\end{equation}
and finally
\begin{align}
  F_{4,1}^{(5c)} & =F_{4,1}^{(5a)}+F_{4,1}^{(5d)}\,,
  \\
  F_{4,1}^{(5d)} & = \frac{8}{nn_1}\frac{1}{V}\sum_{x y k} \langle
  \{\vpix\cdot\left[\vpi''(y)-\vpi'(y)\right]\}\vpi'(x)\cdot\partial_k^2\vpiy
  \nonumber \\
  &
  -\{\vpi'(x)\cdot\left[\vpi''(y)-\vpi'(y)\right]\}\vpix\cdot\partial_k^2\vpiy
  \rangle_0
  \nonumber \\
  & =\frac{8(n_1-1)}{n}\frac{1}{V}\sum_{x y k} \left[\pz^y
    G(x-y-\hat{0})\partial_k^{y2}G(x+\hat{0}-y)
    -\pz^yG(x-y)\partial_k^{y2}G(x-y)\right]
  \nonumber  \\
  & =\frac{8(n_1-1)}{n}\mcF_7\,,
\end{align}
where $\mcF_7$ is given by \eqref{cF7}.

\subsection{\boldmath Computation of $F_5$ up to $\order{g_0^2}$}
\label{AppC5}

\begin{equation}
  F_5=\order{g_0^4}\,, 
\end{equation}
and hence doesn't contribute to the order considered.

\section{\boldmath The $\mathbf{n=2}$ case with lattice regularization}
\label{AppD}

The lattice action with the chemical potential is
\begin{equation}
  \begin{aligned}
    A &= -\frac{1}{g_0^2} \sum_{x \mu} 
    \cos\left(\pmu \Phi(x) - i h \delta_{\mu 0}\right) 
    \\
    &= -\frac{1}{g_0^2} \sum_x  \left[
      \sum_\mu \cos\left(\pmu \Phi(x) \right)
      + i h \sin\left(\pz \Phi(x) \right)
      + \frac12 h^2 \cos\left(\pz \Phi(x) \right)
    \right] + \order{h^3} 
  \end{aligned}
  \label{Aneq2}
\end{equation}

With $\Phi(x) = g_0 \phi(x)$ we have
\begin{equation}
  \left. A \right|_{h=0}= A_0 + g_0^2 A_1 + g_0^4 A_2 + \ldots 
\end{equation}
where
\begin{align}
  A_0 &= \frac12 \sum_{x \mu} (\pmu \phi(x))^2\,,  \\
  A_1 &= -\frac{1}{24} \sum_{x \mu} (\pmu \phi(x))^4\,.
\end{align}
The $h$ dependent part is given by
\begin{equation}
  A_h = -\frac{h^2}{2 g_0^2}V+ i h g_0 B_1
  + h^2 \left( B_{20} + g_0^2 B_{21} + \ldots \right) 
\end{equation}
\begin{align}
  B_1 &=   \frac{1}{6} \sum_{x} (\pz \phi(x))^3\,, \\
  B_{20} &=
  \frac{1}{4} \sum_{x} (\pz \phi(x))^2\,,\\
  B_{21} &= -\frac{1}{48} \sum_{x} (\pz \phi(x))^4\,.
\end{align}
Note that we need the free energy only up to $h^2 g_0^2$; the omitted terms do
not contribute to this order.
\begin{equation}
  \begin{aligned} 
    V f_h &= \langle A_h \rangle -\frac12 \langle A_h^2 \rangle
    + \frac12 \langle A_h \rangle^2  \ldots  \\
    &= -\frac{h^2}{2 g_0^2} V + h^2 \langle B_{20} \rangle_0 + h^2 g_0^2
    \left( \langle B_{21} \rangle_0 - \langle B_{20} A_1 \rangle_0^c + \frac12
      \langle B_1^2 \rangle_0 \right)\,.
  \end{aligned}
\end{equation}
So
\begin{equation}
  \chi = \frac{1}{g_0^2}\left( 1 + g_0^2 R_1 +
    g_0^4 R_2 + \ldots\right)\,,
  \label{fh_n2} 
\end{equation}
with
\begin{equation}
  \begin{aligned}
    R_1 &= -\frac{2}{V}\left\langle B_{20}\right\rangle_0 = -\frac12
    \left\langle
      \frac{1}{V}\sum_x (\pz \phi(x))^2 \right\rangle_0  \\
    &= \frac12 \Box_0 G(0) = -\frac{1}{2V} \psump
    \frac{\hat{p}_0^2}{\hat{p}^2} = -\frac12 I_{11}\,,
  \end{aligned}
\end{equation}
in agreement with \eqref{R1f}, and
\begin{equation}
  \begin{split}
    R_2 & = -\frac{2}{V} \langle B_{21}\rangle_0 +\frac{2}{V} \langle
    B_{20}A_1\rangle_0^c
    -\frac{1}{V} \langle B_1^2\rangle_0   \\
    & = \frac{1}{24}\left\langle \frac{1}{V}\sum_x (\pz \phi(x))^4
    \right\rangle_0 -\frac{1}{48}\left\langle \frac{1}{V}\sum_{x y \mu} (\pz
      \phi(x))^2 (\pmu \phi(y))^4
    \right\rangle_0^c  \\
    & \qquad\qquad -\frac{1}{36} \left\langle \frac{1}{V}\sum_{x y}(\pz
      \phi(x))^3 (\pz \phi(y))^3
    \right\rangle_0  \\
    & = \frac18 \left( \Box_0 G(0)\right)^2 + \frac14 \sum_{x\mu} \Box_0 G(x)
    \Box_\mu G(x) \Box_\mu G(0)
    +\frac16 \sum_x \left( \Box_0 G(x)\right)^3  \\
    & = \frac18 \left( \frac{1}{V} \psump \frac{\hat{p}_0^2}{\hat{p}^2}
    \right)^2 -\frac14 \sum_\mu \left[ \left( \frac{1}{V} \psump
        \frac{\hat{p}_0^2\hat{p}_\mu^2}{\left(\hat{p}^2\right)^2} \right)
      \left( \frac{1}{V} \psum{q} \frac{\hat{q}_\mu^2}{\hat{q}^2} \right)
    \right] - \frac{1}{6} S_{n2}
    \\
    & = \frac18 I_{11}^2 - \frac14 I_{22}I_{11} - \frac{1}{4\ds}(
    I_{11}-I_{22})(I_{00}-I_{11}) - W_{3c}\,.
  \end{split}
\end{equation}
Here
\begin{equation}
  S_{n2} = -\sum_x \left( \Box_0 G(x)\right)^3 = 6 W_{3c}\,.
  \label{Sn2} 
\end{equation}
The last equation follows from the direct comparison with \eqref{W3c}. Also
\begin{equation}
  -\sum_{x\mu} \Box_0 G(x) \Box_\mu G(x) \Box_\mu G(0) 
  = I_{22}I_{11} + \frac{1}{\ds}( I_{11}-I_{22})(I_{00}-I_{11})\,.  
\end{equation}
The result for $R_2$ above agrees with the result in \eqref{R2alatt} for
$n_1=1$.

\end{appendix}

\end{document}